\documentclass[useAMS,usenatbib]{mn2e}
\usepackage{amsmath}
\usepackage[dvips]{graphicx}
\usepackage{listings}
\usepackage{verbatim}
\usepackage{natbib}

\title[Continuum
emission 
around evolved stars at 1.2\,mm]{Continuum
emission 
around AGB stars at 1.2\,mm\thanks{
Based on observations collected at the European Southern Observatory,
La Silla, Chile within program ESO 70.D-0133, 71.D-0072 and 71.D-0600.}
}
\author[S. Dehaes, M.A.T. Groenewegen, L. Decin, S. Hony, G. Raskin and J.A.D.L. Blommaert]{S. Dehaes$^{1}$\thanks{Scientific researcher of the Fund for Scientific Research, Flanders}\thanks{E-mail:
sofie@ster.kuleuven.be}, M.A.T. Groenewegen$^{1}$, L. Decin$^{1}$, S. Hony$^{2}$, G. Raskin$^{1}$ \newauthor and J.A.D.L. Blommaert$^{1}$
\footnotesize\\
$^{1}$ Institute for Astronomy, University of Leuven, Celestijnenlaan 200D, 3001 Leuven, Belgium\\
$^{2}$ Service d'Astrophysique, Bat.609 Orme des Merisiers, CEA Saclay, 91191 Gif-sur-Yvette, France}
\begin{document}

\date{Accepted . Received ; in original form }

\pagerange{\pageref{firstpage}--\pageref{lastpage}} \pubyear{}

\maketitle

\label{firstpage}

\begin{abstract}
It is generally acknowledged that the mass loss of Asymptotic Giant Branch (AGB) stars undergoes variations on different time scales. We constructed models for the dust envelopes for a sample of AGB stars to assess whether mass-loss variations influence the spectral energy distribution (SED). To constrain the variability, extra observations at millimetre wavelengths ($1.2\,$mm) were acquired. From the analysis of the dust models, two indications for the presence of mass-loss variations can be found, being ($1$) a dust temperature at the inner boundary of the dust envelope that is far below the dust condensation temperature and ($2$) an altered density distribution with respect to $\rho(r) \propto r^{-2}$ resulting from a constant mass-loss rate. For 5 out of the 18 studied sources a two-component model of the envelope is required, consisting of an inner region with a constant mass-loss rate and an outer region with a less steep density distribution. For one source an outer region with a steeper density distribution was found. Moreover, in a search for time variability in our data set at 1.2\,mm, we found that WX Psc shows a large relative time variation of $34$\,\% which might partially be caused by variable molecular line emission.
\end{abstract}



\section{Introduction} \label{intro}
When low- and intermediate mass stars enter the AGB phase at the end of their lives, the mass-loss rate exceeds the nuclear burning rate and so dominates the subsequent evolution. In this process, a circumstellar envelope (CSE) of gas and dust is formed which expands at a rate of about $10$\,km\,$\rmn{s}^{-1}$ (see for example \citet{habing}).

The AGB mass loss undergoes variations on different time scales. The mass loss gradually increases over a time scale of several hundred thousands of years to reach a maximum on the tip of the Thermally Pulsing-AGB \citep{habing}. The pulsations of the central star cause variations over several hundreds of days. Strong variations in the mass-loss rate, probably related to thermal pulses occurring every $5$-$10$ $10^4$\,yr \citep{vassiliadis}, can lead to the formation of circumstellar detached shells. \citet{schoier} have recently studied 7 carbon stars with known detached molecular shells. They find that the shells are caused by periods of very intense mass loss ($\sim10^{-5}\,\mathrm{M}_{\odot}\,\mathrm{yr}^{-1}$) and that these shells are running into a previous low mass-loss-rate AGB wind. After the intense mass-losing period, the mass loss decreases to a few $10^{-8}\mathrm{M}_{\odot}$, on a time scale of a few thousand years.

The question then arises if these types of mass-loss variations can be deduced from the SED. To answer this question, we modelled the SED of a carefully selected sample of nearby AGB stars, post-AGB stars and supergiants. For modelling purposes it proved desirable to obtain millimetre observations of the stars in our sample, as there are not many observations available in the literature in this wavelength range. Still, a considerable part of the dust envelope is so cold that it emits its radiation at millimetre wavelengths, making these fluxes necessary to constrain the models for the SEDs. \\

The outline of this article is as follows. First, the obtained fluxes at $1.2\,$mm are presented. Sect.\,\ref{observations} contains information on the sample selection, the observations with SIMBA and the data reduction with {\sc mopsi}. Sect.\,\ref{aperture} is devoted to the aperture photometry.  In Sect.\,\ref{models} the constructed models for the SEDs are presented and the quality of these models and the indications for variable mass loss are discussed. In Sect.\,\ref{variations}, the variability at $1.2$\,mm is analysed on the basis of our observations. The conclusions are summarised in Sect.\,\ref{conclusions}.


\section{Observations and data reduction} \label{observations}

\subsection{Sample selection}
The AGB stars in the sample include M- and C-stars. We have also observed some post-AGB stars and supergiants that are known to have circumstellar shells. To improve the sensitivity to very faint circumstellar emission, the selection was restricted to objects with low mm-background emission based on the \textit{IRAS} 60 and $100\,\umu$m data (typically the CIRR3 flag, which is the total $100\,\mu$m sky surface brightness, $\leq 50\,\rmn{MJy}\,\rmn{sr}^{-1}$) and the sample was further restricted to stars within 1\,kpc (see Table \ref{results}). There are a few exceptions, such as the unique F-type supergiant IRC\,$+10\,420$.

\subsection{Observations}
The data were taken with the 37-channel hexagonal bolometer array SIMBA
installed at the 15-m SEST telescope, during $3$ observation runs: 2002 September 5--10, 2003 May 9--13 and 2003 July 13--15. The channels have a half power beam width (HPBW) of 24\,arcsec and two adjacent channels are
separated by 44\,arcsec. The filter bandpass is centered on 1.2\,mm or
250\,GHz and has a full width half max (FWHM) of 90\,GHz.

SIMBA works without a wobbling secondary mirror; the maps were made with the
fast-scanning technique. The scans were performed in azimuthal direction by
moving the telescope at a speed of $80\,\rmn{arcsec}\,\rmn{s}^{-1}$ with a separation in azimuth of
8\,arcsec. The size of the scans varies between  500\,arcsec x 360\,arcsec and
600\,arcsec x 480\,arcsec. 4 to 8 consecutive scans were made of each source. This
procedure was repeated several times for each source during the observation
runs. Although SIMBA has a hexagonal array design, there are still gaps in the spatial coverage. Because the telescope is set up in horizontal coordinates, the scan direction on the sky changes with the hour angle. By assembling scans taken at different hour angles, the gaps in the spatial coverage are reduced. Of course the assembling of the different scans also increases the signal-to-noise ratio.

\subsection{Data reduction}
The {\sc mopsi} software was used to reduce the data. Basically the procedure described
in the section on {\sc mopsi} in the SIMBA user-manual \citep{handleiding} was followed, but the reduction was further divided into 3 different
steps. In the first step some fundamental operations like despiking, opacity correction and sky-noise
reduction are performed on each of
the individual scans. Also, the 4 to 8 consecutive scans of each source are put together to
form one map. In a second step, the maps made during different nights are
assembled. If the source was detected in the individual maps, the maps were recentered before the assembling. 
Out of this map information is gained about the position of the source,
which is used in the last step to define a proper baseline and to improve the
sky-noise reduction.

For the absolute calibration, scans were made of Uranus, typically at the beginning and/or end of each observing night. The data reduction is basically the same as for the scientific sources, although a few steps were altered because Uranus is a relatively strong and extended source. Only low order baseline fits were used (with an appropriate base range definition) and neither despiking nor sky-noise reduction were carried out, as recommended in the manual. The flux of Uranus at 250\,GHz was estimated with the
subprogram `Planet' in {\sc mopsi}.
Table \ref{tabelvarianties} gives the resulting calibration factors 
for the 3 observation runs.
\begin{figure}
\begin{center}
    \includegraphics[]{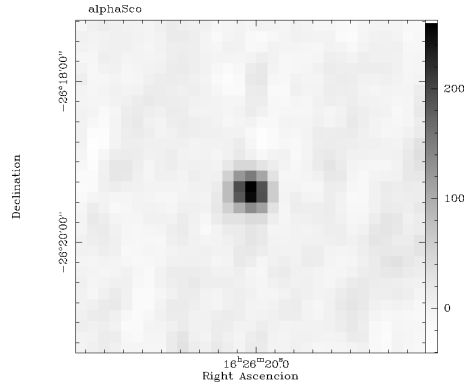}
    \caption{The final map of $\alpha$\,Sco, resulting from the observations of September 2002. The legend is expressed in mJy, the rms-noise on the sky background is $27.7\,\textrm{mJy}$. }
    \label{map}
\end{center}
\end{figure}
As an example, Fig.\,\ref{map} shows the final map of $\alpha$\,Sco resulting from the observations of September 2002.

\begin{table*}
\centering
\begin{minipage}{152mm}
\caption{The first row gives the calibration factors for each of the observation runs. The rest of the table contains the variances of the gaussian profiles for the 3 observation runs and of
the averaged PSF, together with the corresponding FWHM.}
\label{tabelvarianties}
\begin{tabular}{|c|c|c|c|c|}
\hline
         & Sep02  &  May03 & Jul03 & PSF \\ \hline 
calibration factor (to mJy) & $62.0 \pm 2.6$ & $64.5 \pm 2.7$ & $58.6 \pm 2.4$ & \\
 $\sigma_x$  & $10.44 \pm 0.47 $ &  $10.30 \pm 0.45$   & $10.49 \pm 0.73$ &
$10.40 \pm 0.30$ \\
 $\sigma_y$  & $10.55 \pm 0.48 $ &  $9.92  \pm 0.44$   & $11.13 \pm 0.78$ &
$10.43 \pm 0.31$ \\
 $\textrm{FWHM}_x$  & $24.7 \pm 1.2 \,\rmn{arcsec} $ & $24.3  \pm 1.1 \,\rmn{arcsec}$ & $24.7 \pm 1.8 \,\rmn{arcsec}$ &  $24.49 \pm 0.71 \,\rmn{arcsec} $ \\
 $\textrm{FWHM}_y$  & $24.8 \pm 1.2 \,\rmn{arcsec} $ & $23.4  \pm 1.1 \,
\rmn{arcsec} $ & $26.2 \pm 1.9 \,\rmn{arcsec}$  & $24.56 \pm 0.73 \,\rmn{arcsec} $  \\ \hline 
\end{tabular}
\end{minipage}
\end{table*}


\section{Aperture photometry} \label{aperture}
After the data reduction, the fluxes were determined using aperture photometry.
For each source we searched for the `ideal' aperture: the aperture with the
highest corresponding signal-to-noise ratio. For the small apertures a correction is
necessary, because one neglects counts from the source that were detected outside the aperture.
To apply this correction, the point spread function or PSF had to be determined.

The PSF was calculated using the 3 final maps of Uranus from the 3 observation runs. 
We used Uranus and not one of the scientific targets because it is such a
strong source and its angular diameter is well known. Because it is an extended 
source, the maps do not just represent the PSF, but a convolution of the PSF with a uniform disk with diameter equal to the angular diameter of Uranus. As we wanted to model the PSF with a gaussian profile, we fitted a convolution of a uniform disk with an elliptic gaussian profile to each of the $3$ maps and
determined the best fit with a least-squares method. The variances in the x- and
y-direction of the 3 resulting gaussian profiles are given in Table\,\ref{tabelvarianties}, together with the corresponding
FWHM (which is $2\sqrt{2\,\textrm{ln}2}\,\sigma$) in the $2$ directions. The given uncertainties are $1\sigma$ errors.
As can be seen from Table\,1, there is little difference between the variances
of the 3 gaussian fits. The variances of the average PSF were calculated as
weighted averages of $\sigma_x$ and $\sigma_y$ from the 3 fits, which are also
given in Table\,\ref{tabelvarianties}. The corresponding FWHM in the x- and
y-direction of the average PSF are in good agreement with the expected value of
24\,arcsec, which is the HPBW of one channel.

With the help of the average PSF, the scale factors to correct for count losses
with small apertures could be determined (see Table\,\ref{tabelschaalfactor}). 
An aperture of 45\,arcsec  already contains $99.99\%$ of
the flux, so we put the scale factors for apertures $\geq 45$\,arcsec equal to 1.
The scale factors from the 3 individual gaussian fits were also calculated to
determine the rms-errors on the scale factors from the average PSF.

\begin{table}
\centering
\caption{The scale factors to correct for count loss within small apertures.}
\label{tabelschaalfactor}
\begin{tabular}{|c|c|}
\hline
 aperture & scale factor \\ \hline 
 $40\,\rmn{arcsec} $& $0.99946 \pm 0.00014$ \\ 
 $35\,\rmn{arcsec} $& $0.99658 \pm 0.00068$ \\ 
 $30\,\rmn{arcsec} $& $0.9841 \pm 0.0023$ \\ 
 $25\,\rmn{arcsec} $& $0.9438 \pm 0.0054$ \\  
 $20\,\rmn{arcsec} $& $0.8419 \pm 0.0095$ \\  
 $15\,\rmn{arcsec} $& $0.647 \pm 0.012$ \\ \hline 
\end{tabular}
\end{table}
Once these scale factors were known, we continued by determining the ideal aperture and the
corresponding flux for each of the scientific targets.
In the user manual of {\sc mopsi} one can find the commands to determine the number
of counts within a given aperture, as well as the error. We have discovered that the numbers that 
{\sc mopsi} returns are not what we understand to be the number of counts and the error. {\sc mopsi} considers
the number of positive counts in the integration area as the total number of counts and the number
of negative counts as the error. Hence {\sc mopsi} adds the positive noise to the
total number of counts, but neglects the negative noise. In our opinion it is better to take the 
positive counts minus the negative counts as the total number of counts, because
after the data reduction one expects the noise to fluctuate around zero.

To calculate the error on this total number of counts, we assumed that the noise follows a 
Poisson distribution, so that the error equals the square root of the total number of counts. 
When the ideal aperture is less than 45\,arcsec, a scale factor had to be applied to derive the total
number: in that case the error on the scale factor propagates into the error on the counts. 
The total number of counts within the ideal aperture also has to be divided
by the area of the PSF, $\Omega$, being
\[ \Omega = \frac{\rmn{FWHM}_x \ \rmn{FWHM}_y \ \upi}{4 \ \rmn{ln}2}\]

The values of $\rmn{FWHM}_x$ and $\rmn{FWHM}_y$ from the average PSF from 
Table \ref{tabelvarianties} were used. The error on $\Omega$, resulting from the errors on 
$\rmn{FWHM}_x$ and $\rmn{FWHM}_y$, was also accounted for in the error on the total number of 
counts. Finally, we also took the error on the calibration factors
into account (see Table \ref{tabelvarianties}). 

To make sure that we do not exclude any extended emission by using the ideal aperture, we have fitted gaussian profiles to each of the detected sources. We have taken ideal apertures that contain at least $99\,\%$ of the volume underneath those gaussian fits. Besides the gaussian flux profiles of the central sources, no other structures are visible in the maps, so there is no danger of contamination of the flux by other objects, not even for the largest apertures.

The resulting flux and the ideal aperture for each of the science targets are given 
in Table \ref{results}.

\begin{table*}
\centering
 \begin{minipage}{175mm}
  \caption{General information about the science targets, together with the derived flux and the corresponding ideal aperture. In case no detection was made, three times the rms on the sky background is given as an upper limit. A question mark in the column of the distances means that no reliable distance estimate was available. References are given below.}
  \label{results}
  \setlength{\tabcolsep}{1.5mm}
  \begin{tabular}{@{}lcrlcclcc@{}}
  \hline
   Name     & RA    & \multicolumn{1}{c}{Dec}  & spectral type & distance & observation & flux at 1.2\,mm & ideal aperture & rms \\
            &       &      &           &\multicolumn{1}{c}{(kpc)} & run &\multicolumn{1}{c}{(mJy)} & (arcsec) & (mJy) \\
 \hline
AFGL\,$1922$ & $17:04:55.1$ & $-24:40:41$ & C$^a$ & $0.88^e$ &  Sep$02$ &  $93.7 \ \pm \ 6.2$  &  $50$ & $8.0$\\ 
$\alpha$\,Sco & $16:26:20.3$ & $-26:19:22$ & M1.5Iab-b$^b$  & $0.19^f$ & Sep$02$ & $244.4 \, \pm \, 15.1$ &  $35$ & $9.2$\\  
$\alpha$\,Sco & & & & &  Jul$03$ & $237.9 \, \pm \, 14.7$  & $40$ & $8.7$\\
HD\,$44179$ & $06:17:37.0$ & $-10:36:52$ & B8V$^b$ & $0.33^g$ & Sep$02$ & $356.1 \, \pm \, 21.6$ & $40$ & $10.5$ \\
IRAS\,$15194$-$5115$ & $15:19:27.0$ & $-51:15:18$ & C$^a$ & $0.6^h$ & Sep$02$ & $310.6 \, \pm \, 19.0$ & $65$ & $9.8$\\
$o$\,Cet & $02:16:49.1$ & $-03:12:13$ & M7IIIe$^b$ & $0.11^e$ &  Sep$02$ & $180.7 \, \pm \, 11.3$ & $55$ & $8.0$ \\
AFGL\,$3068$ & $23:16:42.4$ & $16:55:10$ & C$^c$ & $0.95^e$ &  Sep$02$ & $429.4 \, \pm \, 25.9$ & $90$ & $9.8$\\
IK\,Tau & $03:50:43.7$ & $11:15:31$ &  M6me$^b$ & $0.26^e$ &  Sep$02$ & $436.0 \, \pm \, 26.3$ & $115$ & $8.6$\\
IRC\,$+10\,420$ & $19:24:26.7$ & $11:15:09$ & F8Ia$^b$ & $4-6^i$ & Sep$02$ & $460.7 \, \pm \, 27.8$ & $100$ & $11.7$\\
VY\,CMa & $07:20:54.8$ & $-25:40:12$ & M3/M4II$^b$ & $0.66^e$ & Sep$02$ & $689.7 \, \pm \, 41.2$ & $60$ & $10.5$\\
WX\,Psc & $01:03:48.1$ & $12:19:51$ & M9$^b$ & $0.54^e$ & Sep$02$ & $328.3 \, \pm \, 20.0$ & $115$ & $6.8$\\
WX\,Psc & & & & & Jul$03$ & $158.1 \, \pm \, 10.0$ & $75$ & $8.7$\\
V$1300$\,Aql & $20:07:47.4$ & $-06:25:11$ & M$^b$ & $0.66^e$ & May$03$ & $103.4\,\pm\,6.9$ & $70$ & $7.0$\\
IRAS\,$08005$-$2356$  & $08:00:32.6$ & $-23:56:16$ & F5e$^b$ & ? & Sep$02$ & $< \, 27.7$ & & $9.2$\\
HD\,$56126$ & $07:13:25.3$ & $10:05:09$ & F5Iab$^b$ & $2.3^j$ & Sep$02$ & $< \, 48.0$ & & $16.0$\\
HD\,$56126$ & & & & & May$03$ & $< \, 51.8$ & & $17.3$ \\
S\,Sct & $18:47:37.2$ & $-07:58:00$ & C6,4(N3)$^d$ & $0.62^e$ & Sep$02$ & $< \, 120.0$ & & $40.0$\\
S\,Sct & & & & & May$03$ & $< \, 32.6$ & & $10.9$\\
HD\,$187885$ & $19:50:00.8$ & $-17:09:38$ &  F2/F3Iab$^b$ & ? & Sep$02$ & $< \, 29.5$ & & $9.8$\\
HD\,$187885$ & & & & & May$03$ & $< \, 23.0$ & & $7.7$\\
HD\,$179821$ & $19:11:25.0$ & $00:02:19$ & G5Ia$^b$ & $1^k$ & Sep$02$ & $< \, 79.4$ & & $26.5$\\
HD\,$179821$ & & & & & May$03$ & $< \, 78.7$ & & $26.2$\\
$89$\,Her & $17:53:24.1$ & $26:03:23$ & F2Ibe$^b$ & $0.98^f$ & May$03$ & $69.5 \, \pm \, 4.8$ & $50$ & $9.6$ \\
AFGL\,$4106$ & $10:21:32.4$ & $-59:16:52$ & K0$^b$ & $3.3^l$ & May$03$ & $96.0 \, \pm \, 6.4$ & $60$ & $9.6$ \\
IRC\,$+10\,216$ & $09:45:14.9$ & $13:30:40$ & C9,5$^d$ & $0.12^e$ & May$03$ & $3764.3\,\pm\,221.4$ & $70$ & $12.2$ \\
IRC\,$+20\,326$ & $17:29:42.5$ & $17:47:27$ & C$^c$ & $0.79^e$ & May$03$ & $8.9\,\pm\,1.1$ & $50$ & $10.9$ \\
IRC\,$+20\,370$ & $18:39:42.0$ & $17:38:10$ & C$^b$ & $0.60^e$ & May$03$ & $58.9\,\pm\,4.2$ & $50$ & $16.0$ \\
R\,Hya & $13:26:58.3$ & $-23:01:25$ & M7IIIe$^b$  & $0.13^e$ & May$03$ & $94.7 \,\pm\,6.3$ & $30$ & $12.8$\\
V\,Hya & $10:49:11.3$ & $-20:59:04$ & C6,3e-C7,5e(N6e)$^d$ & $0.33^e$ &  May$03$ & $99.5 \,\pm\,6.6$ & $35$ & $7.7$\\
W\,Hya & $16:46:12.2$ & $-28:07:07$ & M7IIIe$^b$ & $0.12^e$ &  May$03$ & $280.0 \,\pm\, 17.2$ & $45$ & $7.7$\\
CPD-$56\ 8032$ & $17:04:47.8$ & $-56:50:56$ &  WC$^b$ & $1.35^m$ &  May$03$ & $255.4 \,\pm\, 15.8$ & $100$ & $8.3$\\
IRAS\,$14331$-$6435$ & $14:33:07.9$ & $-64:35:03$ &  B3Iab:e$^b$ & ? &  May$03$ & $94.2 \,\pm\,6.3$ & $85$ & $11.5$\\
U\,Hya & $10:35:05.0$ & $-13:07:26$ & C6.5,3(N2)$^d$ & $0.35^e$ &  May$03$ & $< \, 30.7$ & & $10.2$\\
HR\,$4049$ & $10:15:49.9$ & $-28:44:29$ & B9.5Ib-II$^b$ & $0.67^f$ & May$03$ & $<\,28.8$ & & $10.2$\\
AC\,Her & $18:28:09.0$ & $21:49:53$ &  F4Ibpvar$^b$ & $0.75^n$ &  May$03$ & $<\,107.5$ & & $35.8$\\
$\alpha$\,Her & $17:12:21.9$ & $14:26:45$ & M5Iab$^b$ & $0.12^e$ &  Jul$03$ & $147.5 \,\pm\, 9.4$ & $45$ & $9.3$ \\
R\,Scl & $01:24:40.1$ & $-32:48:07$ & C6,5ea(Np)$^d$ & $0.47^e$ &  Jul$03$ & $61.2 \,\pm \, 4.3$ & $35$ & $4.1$\\
TX\,Psc & $23:43:50.1$ & $03:12:34$ & C7,2(N0)$^d$ & $0.23^e$ &  Jul$03$ & $<\, 26.2$ & & $8.7$\\
\hline
\end{tabular}
\end{minipage}
\flushleft
\footnotesize{
a = \citet{groen},
b = SIMBAD,
c = \citet{kwok},
d = \citet{samus},
e = \citet{loup},
f = \citet{hipparcos},
g = \citet{cohen},
h = \citet{ryde},
i = \citet{jones},
j = \citet{juraa},
k = \citet{josselin},
l = \citet{molster},
m = \citet{deMarco},
n = \citet{jurab}
}

\end{table*}


\section{Modelling of the SED} \label{models}

\subsection{Methodology}\label{method}
For all $23$ sources that were detected with SIMBA, models were constructed for the SED with {\sc dusty} \citep{dusty}. {\sc dusty} is a code that solves the problem of radiation transport in a dusty environment. Its input consists of four categories of parameters: information about the external radiation source, about the dust properties, the density distribution and the overall optical depth of the dust envelope. From this information, the code calculates the dust temperature distribution and the radiation field, by solving a self-consistent equation for the radiative energy density, including dust scattering, absorption and emission. 

For the modelling of our sources, a spherical model was used, in which the external radiation comes from a point source at the centre of the density distribution. The spectral shape of the external radiation is determined by a black body of a given temperature. For the oxygen-rich dust envelopes, we used silicates to model the chemical composition of the CSE. {\sc dusty} contains the optical properties of $3$ types of silicates: hot and cold silicates which are discussed by \citet{ossenkopf} and a silicate with optical constants from \citet{draineandlee} (further abbreviated by SilOw, SilOc and SilDL). For the oxygen-rich envelopes, only models with one type of grain were tested, never a mixture. For the carbon-rich envelopes in our sample, a mixture with varying ratios of amorphous carbon (AMC, with optical constants from \citet{hanner}) and SiC (with optical constants from \citet{pegourie}) was used. The size of the grains was fixed at $0.1\,\umu$m. The inner radius of the dust envelope is fixed by specifying the dust temperature at the inner boundary, {\sc dusty} assumes instantaneous dust formation at this radius. When dealing with objects where the expansion of the envelope is driven by radiation pressure on dust grains, {\sc dusty} can compute the density structure in the envelope by solving the hydrodynamics equations, coupled to the radiative transfer equations. In this case, the only input parameter required is the relative extension of the dust shell in comparison to the inner radius: this parameter was fixed at $5000$ (resulting in a dust temperature at the outer boundary of the dust shell of about $30$\,K, to meet the temperature of the interstellar medium). In these calculations, {\sc dusty} assumes a luminosity of $10^4\,L_{\odot}$, a gas-to-dust mass ratio of $200$ and a dust grain bulk density of $3\,\textrm{g\,cm}^{-3}$. 

To construct the SEDs, photometric data were collected from the astronomical database VizieR and from a report by van der Hucht (2004, private communication) that lists photometric data from the literature between $12\,\mu$m and $6$\,cm for about half of the sources in our sample. To further constrain the model, we made use of the \textit{IRAS} low-resolution spectrometer spectra with wavelengths between $8$ and $23\,\umu$m (obtained from the website http://www.iras.ucalgary.ca/$\sim$volk/getlrs$\_$plot.html). To estimate the interstellar reddening, a program based on the model of \citet{arenou} was used. Every model that was tested, was reddened according to the average interstellar extinction law for the Johnson system from \citet{savage}.

As a starting point, a model with an effective temperature of $2500\,$K and an optical depth at $0.5\,\mu$m of $10$ was taken, with a dust temperature at the inner boundary of the dust envelope of $1000$\,K. For a carbon-rich envelope we started out with a chemical composition with $90\,$\% amorphous carbon and $10$\,\% SiC, for an oxygen-rich envelope SilDL was tried first. \citet{ivezicandelitzur} have made an extensive parameter study from which the influence of each input parameter on the shape and scale of the SED becomes clear. We have done a parameter study ourselves in a smaller parameter space, especially to study the influence of the different parameters on the predicted LRS spectrum, which is not clear from the SEDs presented in \citet{ivezicandelitzur}.\\
\indent Starting from the model described above, the parameters $T_{\rm{eff}}$, $\tau_{0.5}$, $T_1$ and the chemical composition were altered to improve the fit as follows. First the optical depth is adjusted since this parameter has the largest influence on the model out of the four mentioned above. It does not change the overall shape of the SED, a larger optical depth merely shifts the complete SED redwards. Then the chemical composition is determined. For a carbon-rich envelope, the strength of the $11.3\,\mu$m feature increases as the percentage of SiC is increased. Concerning the oxygen rich case, the type of silicate is determined by the relative strength of the $9.7\,\mu$m and $18\,\mu$m features. Since the strength of the features is also strongly dependent on the value of $\tau_{0.5}$, the silicate features can put additional constraint on this parameter as well. The effective temperature of AGB sources typically lies between $2000$ and $3500\,$K. Since this is a small temperature range, the effects on the SED are also relatively small. $T_{\rm{eff}}$ affects the shape of the SED blueward of the peak of the energy distribution and only has a minor scale effect redwards of the peak. $T_{\rm{eff}}$ only plays a role in the SED when $\tau_{0.5} \leqslant 1$, at larger optical depths the external radiation is fully absorbed and has no more influence on the SED. Finally, the dust temperature at the inner boundary of the envelope, $T_1$, is adjusted. $T_1$ has no influence blueward of the SED peak and the shape of the red part of the SED is only affected when $T_1$ becomes smaller than about $500\,$K. The LRS spectrum is also very important in determining the correct value of $T_1$, as the slope of the LRS spectrum clearly decreases with decreasing values of $T_1$. \\
\indent The optical depth is varied in steps of the order of magnitude of $\tau_{0.5}$, e.g. around $\tau_{0.5}=50$, also models with $\tau_{0.5} = 40$ and $60$ are tried. $T_{\rm{eff}}$ is varied in steps of $500\,$K, $T_1$ in steps of $100\,$K and the amount of amorphous carbon in steps of $1\,\%$. Each model is judged by eye, where equal importance is given to the agreement with the SED and the LRS spectrum. The values from the default model were only altered when this brought along a clear improvement of the fit to the observations. \\
\indent Two of the sources in our modelling sample, HD\,$44179$ and CPD-$56$\,$8032$, are said to be surrounded by dusty disks (see respectively \citet{cohenandvanwinckel} and \citet{demarcoandbarlow}), while there is strong evidence that the dust envelopes from VY\,CMa and $89$ Her are not spherically symmetric (see respectively \citep{josselin2000} and \citet{waters}). For these sources no acceptable model could be constructed, which is to be expected as {\sc dusty} assumes spherically symmetric mass loss. For one source, IRAS\,$14331$-$6435$, not enough data were available to constrain a model. For all other $18$ stars, the SEDs are presented in Appendix \ref{ap}, together with the input parameters and the derived parameters from the {\sc dusty} models. To estimate the errors on the input parameters, we varied one parameter while retaining the values from the best model for the other parameters, until the fit became unacceptable. These errors are clearly only estimates, and not formal errors due to interdependencies of the different parameters.

\subsection{Discussion of the SEDs}

\subsubsection{Quality of the models}\label{quality of the models}
In general, the models produced by {\sc dusty} explain the observations quite well, certainly when we keep in mind that a certain scatter of data around the model is to be expected, as our sample stars are variable and the data were taken at different epochs.

There are a few discrepancies that occur in several SEDs. For IRAS\,$15194$-$5115$, R\,Hya, $\alpha$\,Her and IRC+$10\,420$, the observations at visual and/or near infrared wavelengths can not be explained by the models. For IRC\,$+10\,216$, this problem was solved by increasing the size of the dust grains, but this does not work for all problematic cases. When the mass loss is small, the spectrum of the central star still has a large influence on the spectrum we observe, especially at short wavelengths. For these cases the discrepancy can be explained by the fact that the central star was modelled with a black body instead of using a stellar spectrum suitable for the central star. That the influence of the central star is not negligible can also be seen from the LRS spectrum, where sometimes the atmospheric absorption by HCN and $\textrm{C}_2\textrm{H}_2$ can be noticed at $14\,\mu$m.

There are also significant deviations between the observed and modelled submillimetre fluxes. The flux excesses that are seen here, will be discussed in Section \ref{different_expl}. The excess at centimetre wavelengths is discussed separately in Sect. \ref{excess}.

\subsubsection{Excess at (sub)millimetre wavelengths ( $\simeq 100\,\mu$m - $1$\,cm)}\label{different_expl}
For almost all of the SEDs, the model provides a good fit for fluxes at wavelengths shorter than about $100\,\mu$m, but for a few of them the longer wavelength fluxes are not in agreement with any of the models that were tested. This is the case for AFGL\,$1922$, IRAS\,$15194$-$5115$, WX\,Psc, R\,Scl and V$1300$\,Aql. In this section, various possible explanations for the observed (sub)millimetre excess will be discussed.

The excess emission at (sub)millimetre wavelengths can be due to molecular line emission, which is not included in the models. IRC\,$+10\,216$ is known for its rich molecular spectrum. For example, \citet{groen1997} find an on source contribution from molecular lines which amounts to $43\,\%$ within the passband of the instrument used, which lies between $220$ and $280$\,GHz and which has an effective frequency of $243$\,GHz. Nevertheless, the {\sc dusty}-model -- which does not include molecular contributions -- that was fitted to wavelengths shorter than $500\,\mu$m explains the (sub)millimetre data for IRC\,$+10\,216$ well (see Fig.\,\ref{irc+10216}). The contribution from molecular lines is too small to necessitate a different model. Our SEDs are plotted using a log-log scale in units of $\rm{W\,m}^{-2}$ so that the SEDs show the physically relevant energy distribution. We conclude that the molecular contribution does not cause a flux excess on the scale used and therefore does not influence the (sub)millimetre modelling. The molecular contribution to the flux will vary from source to source, depending on a variety of factors, such as the dust-to-gas ratio, the temperature and velocity structure in the envelope, etc. However, it is unlikely that these factors will change the order of magnitude of the molecular contribution. The molecular emission is also proportional to the mass-loss rate, which is the best determined parameter of influence. Since IRC\,$+10\,216$ undergoes substantial mass loss, we do not expect the molecular emission to cause a noticeable excess for any other star in the sample, on the scale used in the SED. Therefore, we deem it unlikely that the molecular contribution will cause a substantial effect in the parameters derived from the SED fitting.\\

The alternative explanation for the observed excesses is an increased amount of dust radiation in the (sub)millimetre range, either (1) because the dust grains radiate more readily at these wavelengths than is assumed in the model or (2) because of the presence of more cold dust than in the model. These effects can be caused by (1) the presence of larger dust grains or (2) by an increased amount of dust farther away from the star. We have explored the first possibility by varying the grain size in our models for three test cases: IRAS\,$15194$-$5115$, R\,Scl and WX\,Psc. Starting from the parameter values of the best fitting model with grain size $0.1\,\mu$m, grain sizes equalling $0.2$, $0.5$, $1$, $5$, $10$ and $50\,\mu$m were tested. Because larger grains not only influence the shape of the SED, but also shift the SED as a whole redwards, the optical depth was decreased to counteract this effect. For IRAS\,$15194$-$5115$, no model could satisfactory reproduce the SED. For the other $2$ sources, it was possible to fit the SED, including the (sub)millimetre fluxes, but these models fail to reproduce the LRS spectrum. As an example: a model with $\tau=5$ and grain size $a=5\,\mu$m fits the SED of WX\,Psc, but the strong features in the LRS spectrum are not reproduced. On the other hand, a model with $\tau=10$ and $a=0.75\,\mu$m reproduces the LRS spectrum, but not the SED. Besides single sized dust grains we also tried models with grain size distributions $n(a) \propto a^{-q}$ for WX\,Psc. We used a standard power index of $3.5$ and a flatter size distribution with power index $2$, grain sizes between $0.01$ and $10\,\mu$m and $\tau=5$ or $10$, but the fits are even less satisfactory than for the single sized grain models. \\
\indent Note that these results imply that a single population of large grains or a grain size distribution including large grains in a spherical configuration can not be invoked to explain the (sub)millimetre excess. It does not exclude the contribution of large grains in general. The presence of an additional shell of large grains can not be modelled with {\sc dusty}. We have used another radiative transfer code, {\sc modust} \citep{bouwman}, to explore the effect of such a shell for WX\,Psc. We have taken the inner shell from the model derived from SED fitting as presented by \citet{decin} and placed an extra shell with grains of size $1$ or $10\,\mu$m at different positions in the envelope ($Y=10, 100, 1000$). Our conclusion is that a shell with larger grains can only make a significant contribution to the SED at (sub)millimetre wavelengths alone, when the shell is placed far from the star. The low dust temperature implies that the shell must contain a relatively large amount of mass, which was modelled in {\sc modust} by increasing the mass-loss rate in the shell. Models with mass-loss variations will be presented in the next paragraph and further discussed in Sect.\,\ref{mass loss variations}. \citet{jurared} have argued that large dust particles formed in a long-lived disk around HD\,$44179$ can produce the observed millimetre continuum, which is in excess of what simple dust models predict. Again, such a non-spherically symmetric configuration in which a region of enhanced density allows large grains to form can not be modelled by {\sc dusty}. \citet{mauron} studied the morphology of AGB envelopes by imaging the circumstellar dust in scattered light at optical wavelengths. For V$1300$\,Aql they found no evidence of special structures in the envelope, although the envelope itself has an elliptical and not a spherical shape. WX\,Psc seems to have a spherically symmetric extended envelope, but a strong axial symmetry close to the star. R\,Scl has a spherically symmetric detached shell (see \citet{gonzalez}), hence this shell is probably the cause of the (sub)millimetre continuum and there is no need to invoke the presence of large grains. \citet{feast} argue that IRAS\,$15194$-$5115$ has a quasi spherical dust shell, with small scale irregularities due to the ejection of dust clouds. No information was found on the geometry of the CSE for AFGL\,$1922$. Hence on the basis of observations of the structure of their envelopes, it seems that a nonspherical distribution of large grains is an unlikely scenario at least for V$1300$\,Aql and R\,Scl.  \\

To test whether the observed excess is due to dust far from the star, we increase the dust density there in the model. The density distribution as calculated from hydrodynamics assumes a constant mass-loss rate and can be approximated by a distribution $\rho(r) \propto r^{-2}$. Besides a calculation from hydrodynamics equations, {\sc dusty} also offers the possibility to define the density structure with the use of power laws. We tested models with $\rho$ as a function of the relative radius $y$ (scaled to the inner radius of the dust envelope) defined as
\begin{displaymath}
\rho(y) \propto \left\{ \begin{array}{llllll} 
                        y^{-2} & \ \ \ 1 & \leq & y & \leq & y_1\\
			y^{-p} & \ \ \ y_1 & \leq & y & \leq & Y\\
			\end{array}
		\right.
\end{displaymath}
with p equal to $1.5,\,1$\,or\,$0.5$ and $y_1$ equal to $10,\,100$\,or\,$1000$, i.e. models with a higher mass loss in the past. For AFGL\,$1922$, IRAS\,$15194$-$5115$, R\,Scl, WX\,Psc and V$1300$\,Aql, such a density distribution improved the fit, both for the SED and for the LRS spectrum. The fact that the density distribution from the hydrodynamics equations is not adequate is an indication for non-constant mass loss. This will be further discussed in Section \ref{mass loss variations}. 

We want to mention IRC\,+$20\,326$ here as well. For this source, the hydrodynamical models overestimated the observational fluxes. Models using smaller dust grains ($a=0.01$ or $0.001\,\mu$m), could not explain the shape of the SED nor the LRS spectrum. The best fit is obtained by using a model with a steeper density distribution $\rho \propto r^{-3}$ in the outer envelope, consistent with a lower mass loss in the past.

The altered density distribution improves the fit to the data, but it can not explain the feature around $1$\,mm for AFGL\,$1922$ (see Fig.\,\ref{afgl1922}). This feature might also appear in the SED of IRAS\,$15194$-$5115$ (Fig.\,\ref{iras15194}), but the fewer observations around this wavelength make it difficult to tell. A search for the origin of this feature is beyond the scope of this article, but molecular emission can be excluded due to the size of the excess.

\subsubsection{Excess at centimetre wavelengths }\label{excess}

On the basis of the agreement between model and observations at centimetre wavelengths, our sample stars can be divided into two groups. For group\,I, which is the largest group, it was possible to construct models which provide a good overall fit. W\,Hya is a perfect example of this (see Fig. \ref{WHya}). In group\,II, the fit at optical and IR wavelengths is very good, but the models can not explain the observations at radio wavelengths. $\alpha$\,Her is representative for this group (see Fig. \ref{alfaHer}). For $\alpha$\,Her the fit is reasonably good up until 1.2\,mm, the first observations at longer wavelengths are in the centimetre range and these fluxes are underestimated by the model. Although we are only talking about two observations, we do think that this phenomenon is real and not due to for example a bad choice of model parameters. Firstly, this effect can be seen in several SEDs. We see similar excesses for IRC\,$+20\,370$, WX\,Psc, $\alpha$\,Sco and V\,Hya. For all of these stars there are only one or two observations at centimetre wavelengths. For $o$\,Cet and IRC\,$+10\,216$ the excess is less obvious, but in these cases there are $4$, respectively $8$ fluxes that lie systematically higher than the model predictions. 
Secondly, different models over a wide parameter range were tested, but non could provide a better fit than the one shown. We investigated the influence of the different parameters in {\sc dusty} on the resulting models and we found that only the density distribution of the dust can make a significant difference in flux at millimetre and centimetre wavelengths, without affecting the flux at visual and infrared wavelengths. Models with density distributions defined by power laws as explained in Section \ref{different_expl} were tested; this sometimes improved the fit, but it could never fully explain the observed excess. \\

In the literature, there are several articles that comment on a similar flux excess for evolved stars. In \citet{skinner}, models are constructed for the outer layers of the atmosphere and the dust envelope of the supergiant $\alpha$\,Ori. A composite model of photosphere plus chromosphere plus dust emission can explain the flux below $1$\,mm wavelength, but to explain the excess at millimetre-centimetre wavelengths, yet another component (being a partially ionised wind in combination with areas of infalling ionised material) has to contribute to the flux. In \citet{knappb}, radio continuum observations from a sample of 21 evolved stars with high mass-loss rate and an extended CSE are studied, in the hope of detecting new planetary nebulae. For $4$ stars, excess emission is detected that is too weak to be attributed to a compact planetary nebula. The authors suggested a partially ionised chromosphere as the source of this emission. 

Hence one possible explanation of the excesses we observe is the presence of a chromosphere. $\alpha$\,Sco, which is an early M-type supergiant like $\alpha$\,Ori, is known to have a chromosphere, this is also the case for $\alpha$\,Her. The presence of chromospheres in supergiants is widely accepted, however, it is not yet clear if all AGB stars have chromospheres as well. \citet{drake} performed a radio-continuum survey of cool M- and C-giants in a search for chromospheric activity. They concluded that at least in some red giants, chromospheres are either absent or they contain so little ionised gas that their optical depth at centimetre wavelengths is less than unity. $o$\,Cet, WX\,Psc, V\,Hya and IRC\,+$20\,370$ all fall in this category of cool M- and C-giants. Sometimes these stars are said to have a chromosphere in the literature, but without convincing arguments. 

In \citet{sahai} several mechanisms are discussed to explain the centimetre emission from IRC\,$+10\,216$. They concluded that a simple black body in combination with dust emission from amorphous carbon is not sufficient to explain the observed fluxes. The most probable cause of the excess is an increase in dust emission at the relevant wavelengths, which can be justified if long carbon molecules (such as PAHs) are present in the dust envelope. In \citet{groenewegen}, where a model for the circumstellar dust shell of IRC\,+$10\,216$ is presented, the dust opacity also had to be increased for wavelengths larger than $1$\,mm. There is free-free emission present as well, this emission is negligible for wavelengths below $5$\,mm, but does make a significant contribution to the flux at centimetre wavelengths. In this case the free-free emission is not due to a chromosphere, but to a small partially ionised region with temperatures close to that of the photosphere.\\

In summary, the most probable explanation for IRC\,+$10\,216$ is an increase in dust emission at centimetre wavelengths as proposed by \citet{sahai}, which could be caused by the presence of long carbon molecules (such as PAHs) in the CSE. $\alpha$\,Sco and $\alpha$\,Her are known to posses a chromosphere, which could explain the excess emission.  For $o$\,Cet, WX\,Psc, V\,Hya and IRC\,+$20\,370$ there is no clear evidence for the presence of a chromosphere. For V\,Hya and IRC\,$+20\,370$, which have a carbon-rich CSE, the same mechanism as for IRC\,$+10\,216$ might explain the observed excess.

\subsubsection{Mass-loss variations} \label{mass loss variations}
As already mentioned in Section \ref{quality of the models}, we propose a less steep density distribution than the one calculated from hydrodynamics to explain the (sub)millimetre fluxes for AFGL\,$1922$, IRAS\,$15194$-$5115$, R\,Scl, WX\,Psc and V$1300$\,Aql. \citet{hofmann} made an elaborate study of WX\,Psc and they also found that a two component model with an inner constant mass loss region ($\rho \propto r^{-2}$) and an outer region with $\rho \propto r^{-1.5}$ gives a better match to their observations than a one component model with a uniform outflow. The necessity of a density distribution defined by power laws can be seen as an indication for variable mass loss. All five power laws that were fitted indicate that the mass loss has decreased with time. 

For IRC\,+$20\,326$, a steeper density distribution $\rho \propto r^{-3}$ in the outer envelope was necessary to explain the data. This indicates that the mass loss has increased with time. There is one source, AFGL\,$4106$, where the outer radius of the envelope had to be decreased so that the lowest dust temperature amounts to $42$\,K instead of $30$\,K or lower for the other sources. This might indicate that the past mass-loss rate was so low that we can not detect dust at lower temperatures. We were unable to model this source with a steeper density distribution in the outer envelope instead of the density-cutoff.

There are several mechanisms that might cause mass-loss-rate modulations in AGB stars, each one related to quite distinct time scales. Modulations induced by stellar pulsations would display a period of a few hundred days. Thermal pulses occur only once in ten thousand to hundred thousand years. A third timescale of 200 to 1000 years is derived from the separation between the nested shells seen around IRC\,+$10\,216$ \citep{mauronandhuggins} and around several (proto-)planetary nebulae. \citet{simis} suggest that the interplay between grain nucleation and wind dynamics leads to quasi-periodic changes in the mass-loss rate on this time scale. \citet{soker} proposes a solar-like magnetic activity cycle as a mechanism to produce the observed geometry of the CSE surrounding these objects.

From the distance where we find the density profile changes, i.e. the place in our model where the break-point occurs, we can get a rough idea of the duration of the constant mass-loss phase. Translating the break-points into time scales using a typical outflow velocity of $10$ km\,s$^{-1}$, we find dynamical ages between 700 $-$ 20\,000~yr, with an average of 3500~yr, taken into account only those sources that show evidence of mass-loss variability. Of course, if the ``constant'' mass-loss sources are taking into account this average dynamical age will increase substantially since they do not trace variability out to 
$\sim$80\,000~yr, i.e., the outer radius in our modelling. These time scales exclude stellar pulsations as a possible cause of the mass-loss-rate changes detected in the SED. The lower limit of the kinematic age is still in agreement with the third timescale mentioned above. However, the concentric shells seen in IRC\,+$10\,216$ by \citet{mauronandhuggins} that have lead to this timescale, do not leave their mark in the SED (see Fig.\,\ref{irc+10216}). This leaves thermal pulses as the most plausible explanation for the observed mass-loss-rate modulations.

We want to stress here the importance of (sub)millimetre observations to determine the correct density distribution. WX\,Psc is a clear example. Up to $100\,\mu$m, a density distribution determined from hydrodynamics provides a good fit, but the observations at $400\,\mu$m and $1.2\,$mm were clearly underestimated by such a model. On the basis of these observations, a power law density distribution was determined that provided a better overall fit. Measurements at centimetre wavelengths can not be used in determining the correct density distribution, because of the excesses seen at these wavelengths, as discussed in Sect.\,\ref{excess}.\\

A second type of variation in mass loss is indicated by a low dust temperature at the inner boundary of the dust envelope. Under normal circumstances one expects dust to form once the dust condensation temperature is reached. This dust condensation temperature differs for different species of dust, but for silicates and amorphous carbon $1000\,$K can be seen as a typical value. When a dust temperature of $150\,$K is necessary to model the SED, as is the case for AFGL\,$4106$, this means that the dust that dominates the SED has temperatures far below the dust condensation temperature, so it is located at very large distances from the central star. This implies that the past mass-loss rate was far greater then the present mass-loss rate. For AFGL\,$4106$, which is a post-AGB star, the mass loss has stopped and we are dealing with a detached shell.\\

All in all, it is not possible to derive detailed information on variations in the mass-loss rate from the SED alone. First of all, {\sc dusty} contains a variety of input parameters and the models presented here are probably degenerate. To determine the correct model for a given star, as many input parameters as possible should be determined in an independent way. With interferometric techniques for example, the inner radius of the dust envelope can be determined, and thus the temperature at the inner edge. Secondly, the physics in {\sc dusty} can be further improved. In the current version there is no way to model two dust shells with each a constant, but different mass-loss rate. Of course, this would even further increase the level of degeneracy of the models. We can conclude that the SED can indicate variations in mass-loss rate, but it should be seen as a basis for further analysis of the object and not as a final stage of the study.  


\section{Variability at 1.2\,\lowercase{mm}} \label{variations}

Several sources were observed during $2$ different observation runs. Only $2$ of those
sources, being $\alpha$\,Sco and WX\,Psc, were strong enough to produce a flux above the detection limit of SIMBA. 

From Table \ref{results} it can be seen that the 2 fluxes of $\alpha$\,Sco agree with each other within the errors. One does not expect to see much variation, as the $2$ observations are separated by $310$\,d, while the period of $\alpha$\,Sco is $350$\,days \citep{percy}.

WX\,Psc, a Mira variable with a period in the Johnson \textit{K} band of $660$\,d \citep{samus}, does show a clear flux variation. If we take our $2$ observations to be coincident with minimum and maximum light, we find an amplitude of $85.1\,$mJy and a relative variation (the ratio between the amplitude and the average flux at $1.2$\,mm) of $34\,\%$. This constitutes a lower limit to the true amplitude at $1.2$\,mm. In Table \,\ref{fluxenomont} observations at $1.3$\,mm are listed that were obtained with the SEST bolometer between March 1992 and March 1994 (Omont, private communication). If we shift these measurements to $1.2\,$mm, correcting the flux with a power law that was fitted to our model of WX\,Psc between $1.1$ and $1.4\,$mm
\[ F_{\nu\,at\,1.2\,\textrm{mm}} = F_{\nu}\left(\frac{\lambda[\textrm{mm}]}{1.2}\right)^{4.3} \]
we find values of $86.1$ and $80.4$\,mJy. This points towards an even larger amplitude and relative variation.
\begin{table}
\centering
\caption{Fluxes at $1.3$\,mm from WX\,Psc and IRC\,+$10\,216$ that were obtained between March 1992 and March 1994 with the SEST bolometer (Omont, private communication). No errors were derived.}
\label{fluxenomont}
\begin{tabular}{|l|c|c|}
\hline
 Name          & flux   & epoch     \\ 
               & (mJy)  & (JD)      \\ \hline 
 WX\,Psc       & $61$   & $2448891$ \\
               & $57$   & $2448941$ \\
 IRC\,+$10\,216$ & $2218$ & $2448695$ \\
               & $2363$ & $2448764$ \\
	       & $2073$ & $2448942$ \\
	       & $1540$ & $2449070$ \\
	       & $1327$ & $2449164$ \\
	       & $2417$ & $2449315$ \\
	       & $1965$ & $2449440$ \\ \hline 
\end{tabular}
\end{table}
In the literature, there is not much information to be found on flux variations at such long wavelengths. We mention here a result on the variability of IRC\,+$10\,216$, because it is also a Mira variable. In \citet{sandell} a figure is shown of the light curve at $1.1\,$mm. We searched for a period in these data using Fourier analysis and we found $655.3\,$d as a period from the Lomb-Scargle periodogram. A fit with the following sine function 
\[ S(\rmn{mJy}) = 2329.9 + 362.1\,\textrm{sin}\left( 2\,\upi\, \frac{JD\,-\,2447301}{655}\,\right) \]
has a variance reduction of $79\,\%$. Here, the relative variation is $16\,$\,\%. From these values we have to conclude that the relative variation found for WX\,Psc at $1.2\,$mm is rather large. 
We also analysed the $1.3$\,mm data from IRC\,+$10\,216$ that are listed in Table\,\ref{fluxenomont}. We fitted a sine function with the period found at $1.1$\,mm, this gives a variance reduction of $77\,\%$ and a relative variation of $25\,\%$, which is about $10\,\%$ larger than the relative variation at $1.1$\,mm. Molecular line emission could explain this difference, as it contributes significantly to the flux at $1.2$-$1.3$\,mm: as already mentioned in Sect.\,\ref{different_expl}, \citet{groen1997} find an on source contribution from molecular lines of $43\,\%$. If the molecular line emission is variable, the observed variations are a superposition of the variations in the dust emission and in the molecular line emission. The larger relative amplitudes at $1.3$\,mm could be caused by variability in the molecular line emission. As the contribution to the flux from molecular lines within the $1.2$\,mm SIMBA passband is significant, this could also explain the large relative amplitude for WX\,Psc at $1.2$\,mm.


\section{Conclusions} \label{conclusions}
AGB mass loss undergoes variations on different time scales. In this article, we have
addressed the question whether these variations also leave their signature in the SED. In order to find the answer, models for the SEDs were presented for a sample which contained mostly AGB stars. As there are few observations available in the millimetre wavelength area, the selected targets were observed at $1.2\,$mm with SIMBA. 

The presented models were constructed with {\sc dusty}. Although there are some discrepancies between observations and model that occur in several SEDs, there is a good overall agreement between models and observations. Five out of $18$ sources show a flux excess at (sub)millimetre wavelengths. A modified density distribution compared to a constant mass-loss rate, expressed in terms of power laws and indicative of mass-loss variability, improves the agreement for all five sources. This is why we propose that these five stars have experienced such strong variations in their mass loss that this is detectable in their SED in the form of a (sub)millimetre excess. The density distributions that were finally used to model these sources, show that the mass-loss rate was higher in the past for all five stars. The presence of large grains in areas with increased density due to asymmetries in the dust envelope can not be ruled out as an alternative explanation, although observations of R\,Scl and V$1300$\,Aql show no such asymmetries in the envelopes of these stars. One other source, IRC\,$+20326$, required a density distribution which implies a lower mass-loss rate in the past. Besides these altered density distributions, a second indication for variations in the mass loss can be found in a dust temperature at the inner boundary of the dust envelope that lies far below the dust condensation temperature, which indicates the presence of a detached shell. We conclude that the SED can contain clear indications for variable mass loss, but that it is not possible to derive detailed information on the basis of the SED alone. This is due to the uncertainties on the many free parameters in the {\sc dusty} models and because there are not many options in {\sc dusty} to adequately model a variable mass loss.

We also analysed the time variability in our $1.2\,$mm data for $\alpha\,$Sco and WX\,Psc, which were observed during two different observation runs. We found that WX\,Psc shows a relative variation of at least $34\,\%$, which is rather large when compared to the relative variation of $16\,\%$ for IRC\,+$10\,216$ at $1.1$\,mm. We propose that a contribution to the flux from variable molecular line emission might explain the large relative variation of WX\,Psc at $1.2\,$mm.


\section*{Acknowledgements}

SD and LD acknowledge financial support from the Fund for Scientific Research - Flanders (Belgium), SH acknowledges financial support from the Interuniversity Attraction Pole of the Belgian Federal Science Policy P5/36.


\appendix

\section{dust models}\label{ap}

\begin{figure}
\begin{center}
    \includegraphics[]{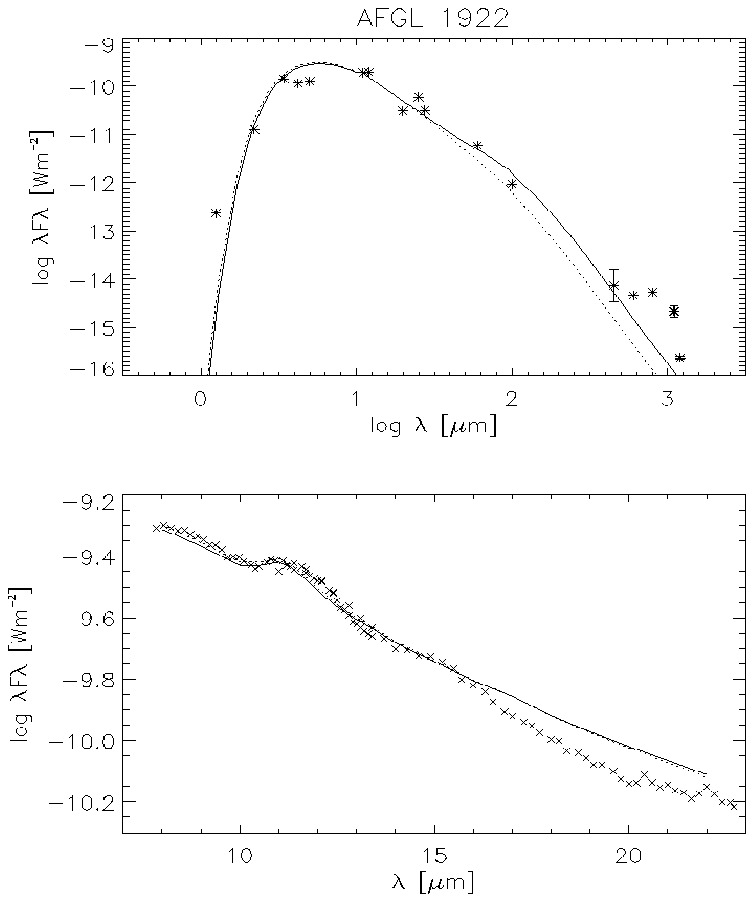}
    \caption{$T_{1} = 1200\,$K, AMC $= 93\,\%$, $\tau_{0.5} = 50$, $T_{\rm{eff}} = 2500\,$K, $\rho(y) \propto y^{-2}$ between $Y=1$ and $Y=100$,  $\rho(y) \propto y^{-1.5}$ between $Y=100$ and $Y=5000$. The upper figure shows the photometric data (asterisk) and the model (full line). Error-bars are shown, but most of the time they fall within the symbols. A reversed triangle represents an upper limit. The lower figure shows the LRS spectrum (plus signs) and the model (full line). As a comparison, the dotted line shows the corresponding hydrodynamical model.}
    \label{afgl1922}
\end{center}
\end{figure}

\begin{figure}
\begin{center}
    \includegraphics[]{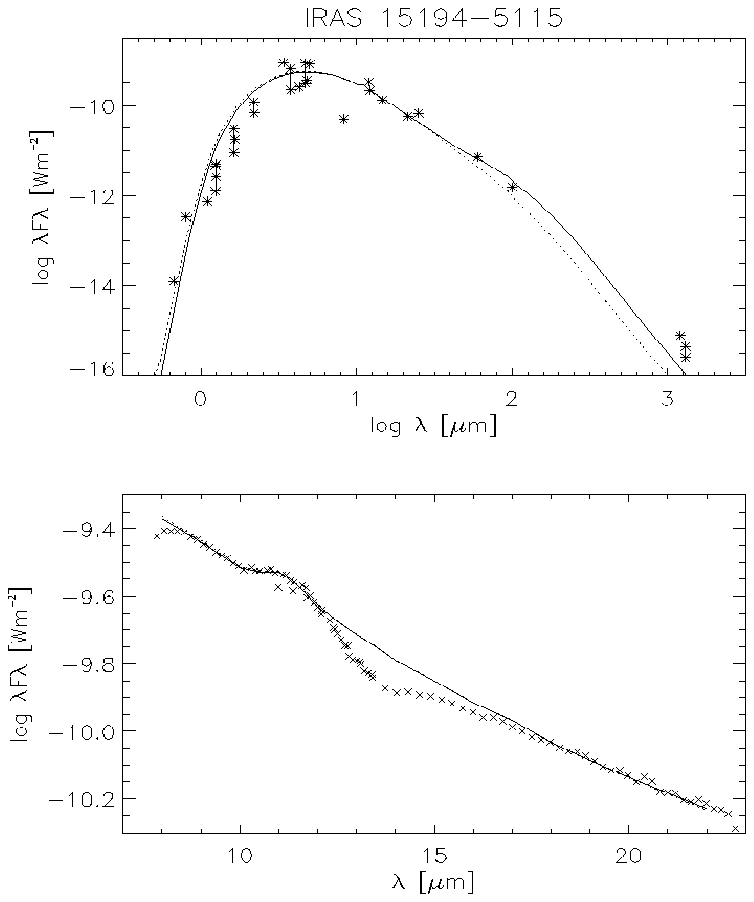}
    \caption{$T_{1} = 900\,$K, AMC $= 96\,\%$, $\tau_{0.5} = 15$, $T_{\rm{eff}} = 2500\,$K, $\rho(y) \propto y^{-2}$ between $Y=1$ and $Y=100$,  $\rho(y) \propto y^{-1.5}$ between $Y=100$ and $Y=5000$. }
    \label{iras15194}
\end{center}
\end{figure}

\begin{figure}
\begin{center}
    \includegraphics[]{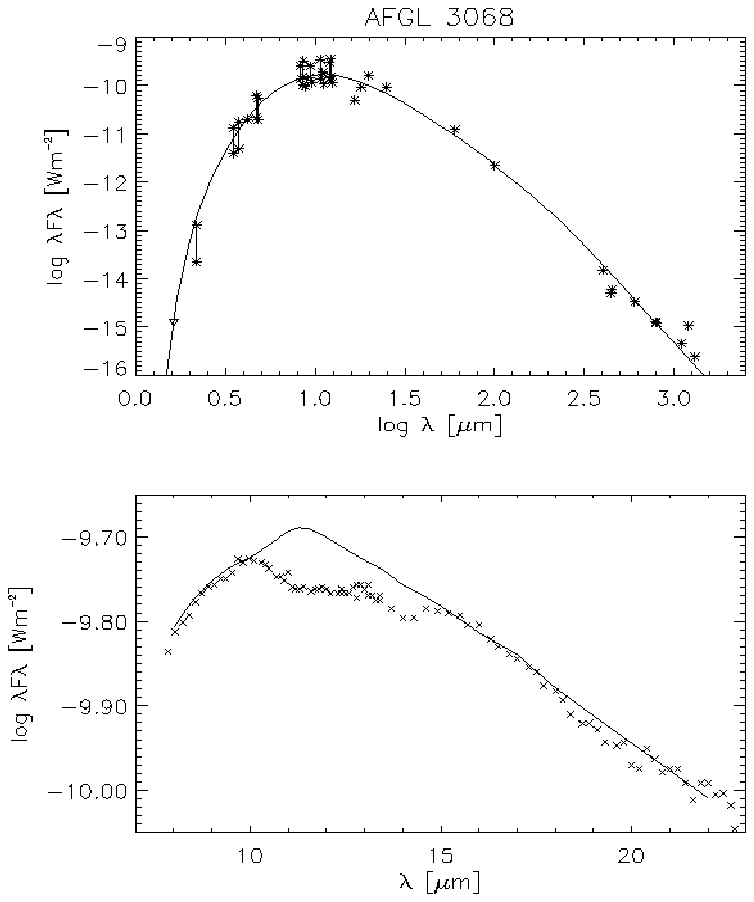}
    \caption{ $T_{1} = 600\,$K, AMC $= 93\,\%$, $\tau_{0.5} = 80$, $T_{\rm{eff}} = 2500\,$K.}
    \label{afgl3068}
\end{center}
\end{figure}

\begin{figure}
\begin{center}
    \includegraphics[]{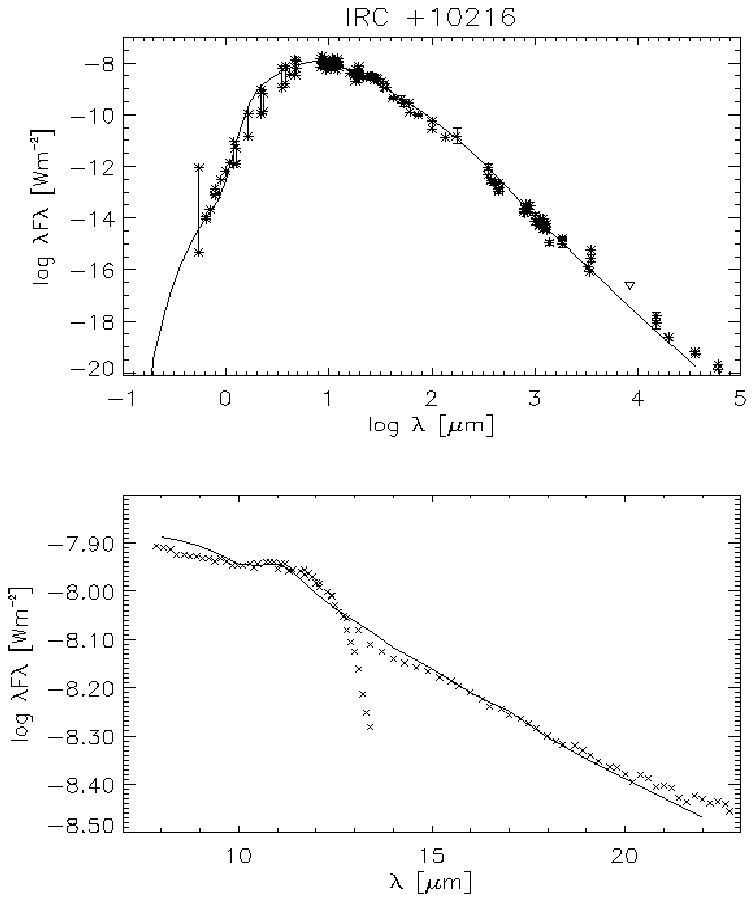}
    \caption{$T_{1} = 600\,$K, AMC $= 97\,\%$, $\tau_{0.5} = 15$, $T_{\rm{eff}} = 2500\,$K, $a= 0.15\,\mu$m.}
    \label{irc+10216}
\end{center}
\end{figure}

\begin{figure}
\begin{center}
    \includegraphics[]{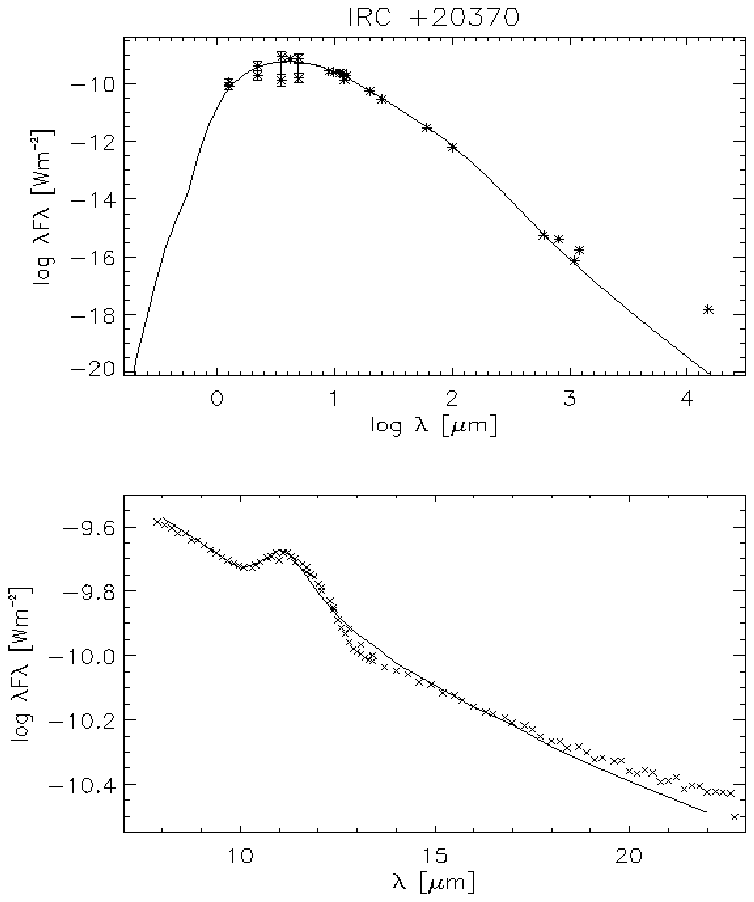}
    \caption{$T_{1} = 900\,$K, AMC $= 90\,\%$, $\tau_{0.5} = 10$, $T_{\rm{eff}} = 2500\,$K.}
    \label{irc+20370}
\end{center}
\end{figure}

\begin{figure}
\begin{center}
    \includegraphics[]{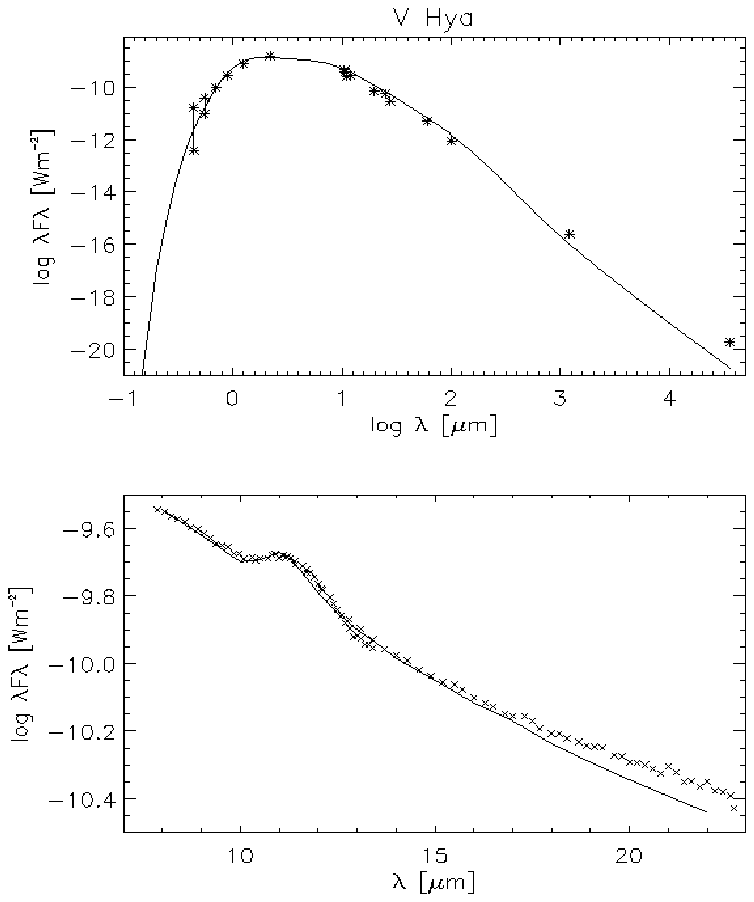}
    \caption{$T_{1} = 700\,$K, AMC $= 93\,\%$, $\tau_{0.5} = 3$, $T_{\rm{eff}} = 2500\,$K.}
    \label{vhya}
\end{center}
\end{figure}

\begin{figure}
\begin{center}
    \includegraphics[]{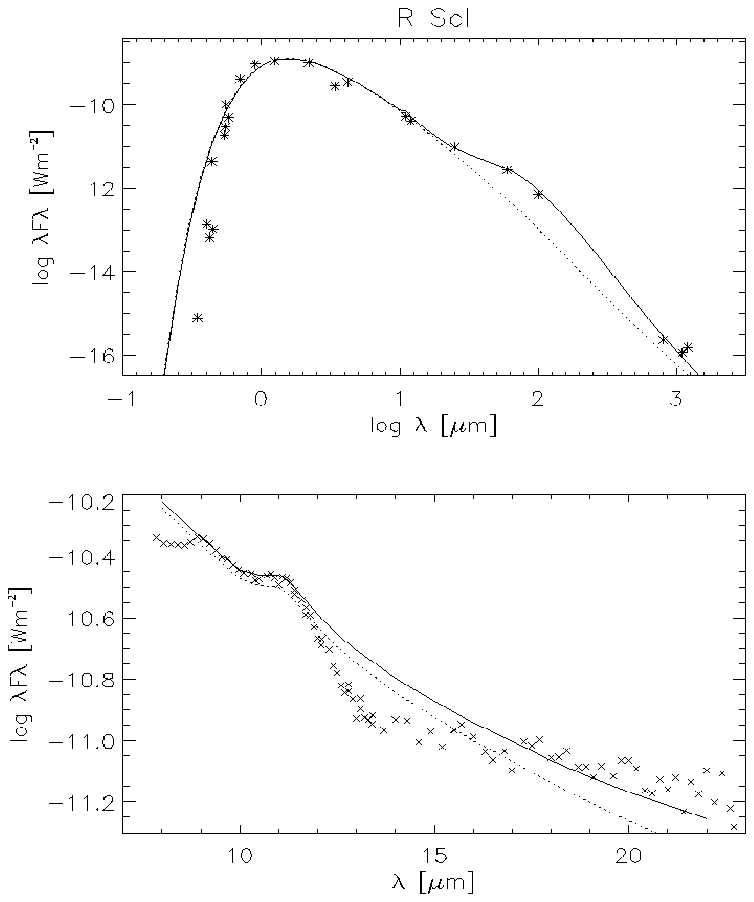}
    \caption{$T_{1} = 1000\,$K, AMC $= 90\,\%$, $\tau_{0.5} = 0.5$, $T_{\rm{eff}} = 2500\,$K, $\rho(y) \propto y^{-2}$ between $Y=1$ and $Y=100$,  $\rho(y) \propto y^{-1}$ between $Y=100$ and $Y=5000$.}
    \label{rscl}
\end{center}
\end{figure}

\begin{figure}
\begin{center}
    \includegraphics[]{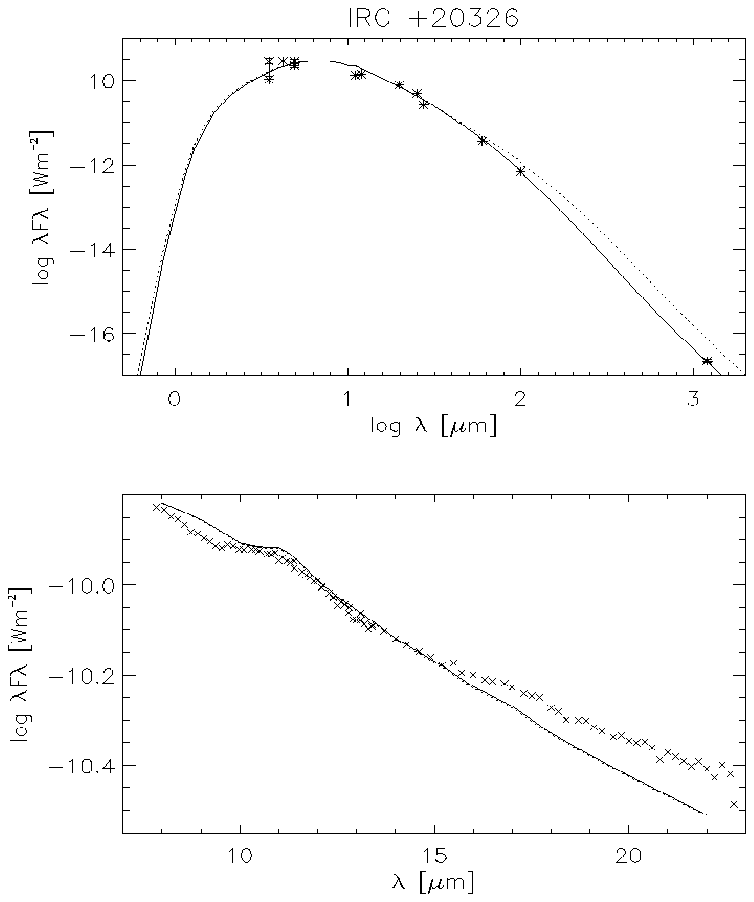}
    \caption{$T_{1} = 700\,$K, AMC $= 97\,\%$, $\tau_{0.5} = 20$, $T_{\rm{eff}} = 2500\,$K, $\rho(y) \propto y^{-2}$ between $Y=1$ and $Y=100$,  $\rho(y) \propto y^{-3}$ between $Y=100$ and $Y=5000$.}
    \label{irc+20326}
\end{center}
\end{figure}

\begin{figure}
\begin{center}
    \includegraphics[]{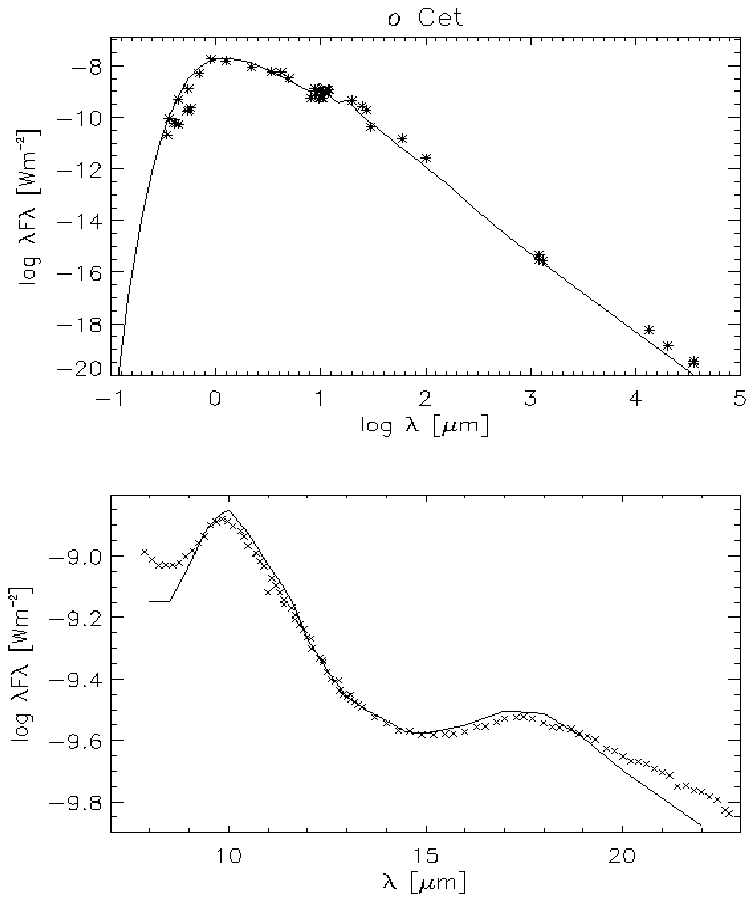}
    \caption{$T_{1} = 1000\,$K, SilOc, $\tau_{0.5} = 1$, $T_{\rm{eff}} = 3000\,$K. }
    \label{ocet}
\end{center}
\end{figure}

\begin{figure}
\begin{center}
    \includegraphics[]{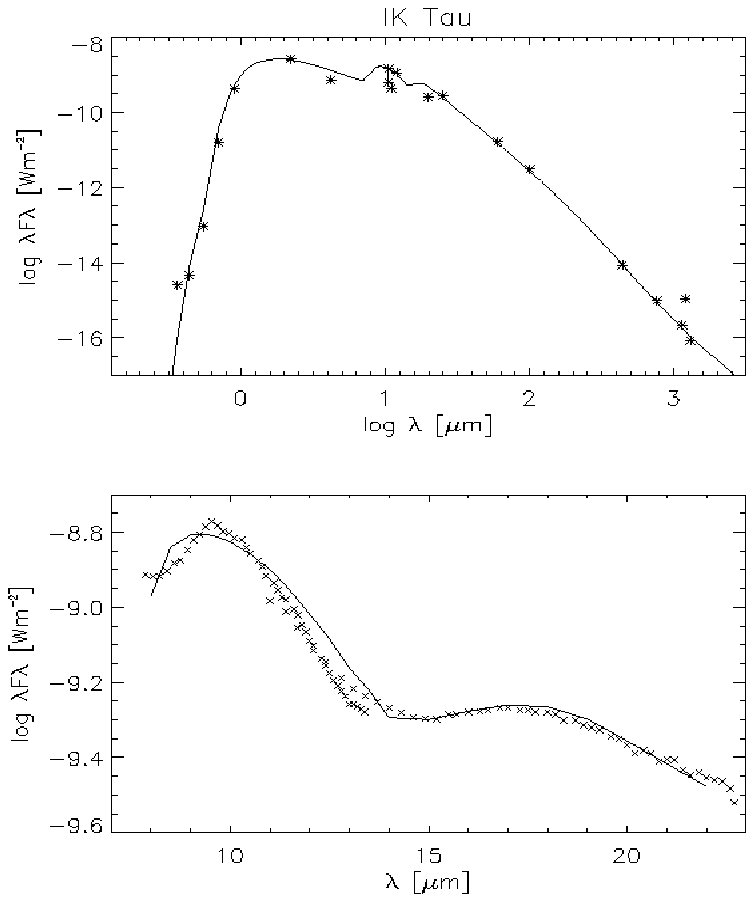}
    \caption{$T_{1} = 1000\,$K, SilDL, $\tau_{0.5} = 20$, $T_{\rm{eff}} = 2500\,$K, $a= 0.15\,\mu$m.}
    \label{IKTau}
\end{center}
\end{figure}

\begin{figure}
\begin{center}
    \includegraphics[]{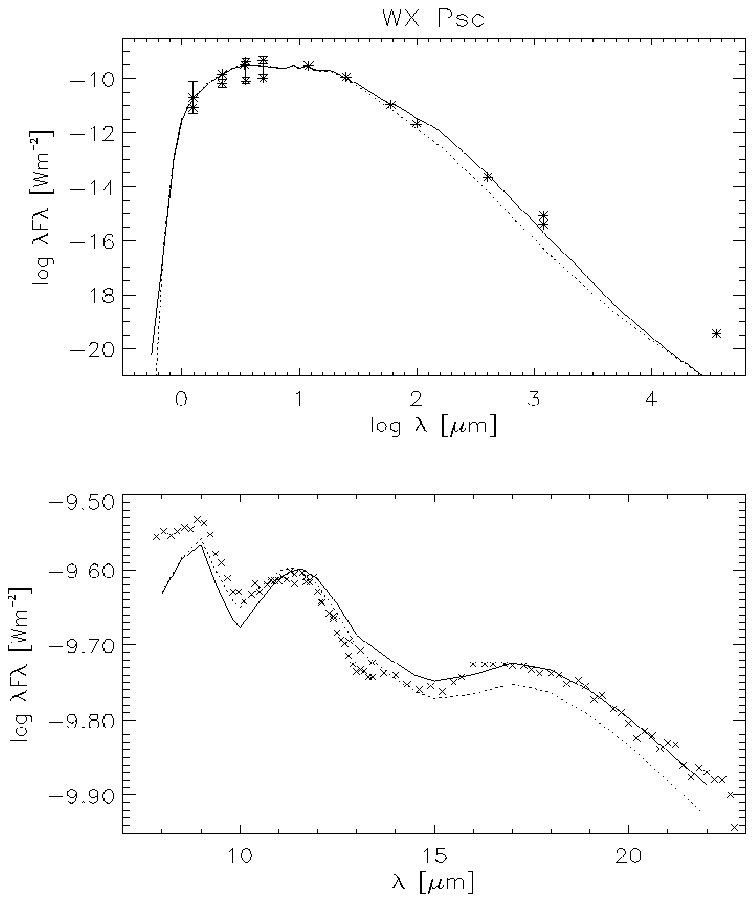}
    \caption{$T_{1} = 1000\,$K, SilOw, $\tau_{0.5} = 70$, $T_{\rm{eff}} = 2500\,$K, $\rho(y) \propto y^{-2}$ between $Y=1$ and $Y=100$,  $\rho(y) \propto y^{-1.5}$ between $Y=100$ and $Y=5000$.}
    \label{WXPSc}
\end{center}
\end{figure}

\begin{figure}
\begin{center}
    \includegraphics[]{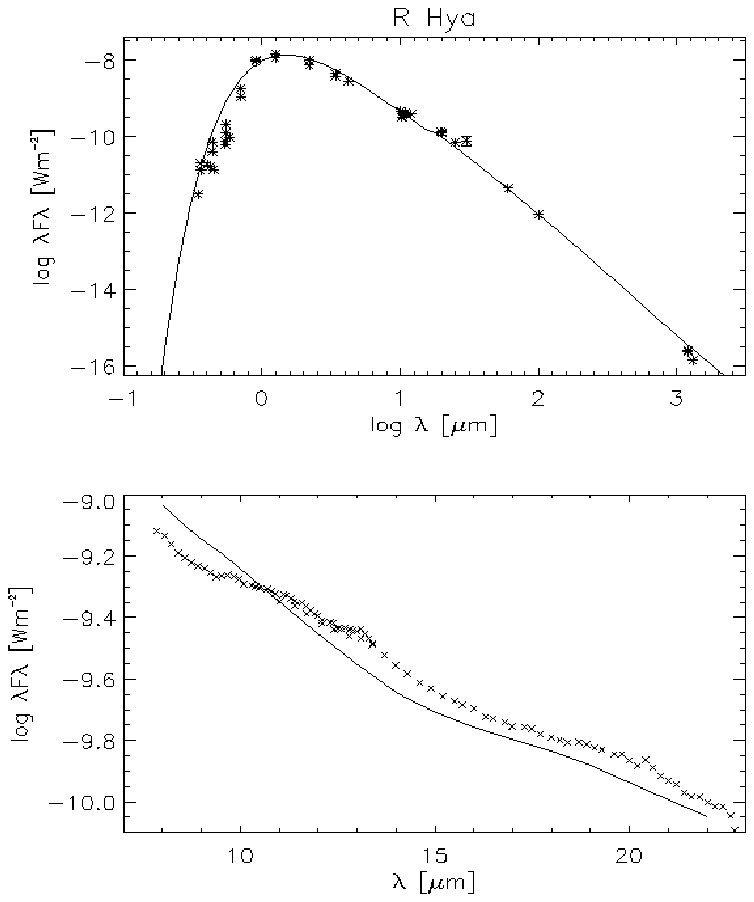}
    \caption{$T_{1} = 300\,$K, SilDL, $\tau_{0.5} = 0.1$, $T_{\rm{eff}} = 2500\,$K.}
    \label{RHya}
\end{center}
\end{figure}

\begin{figure}
\begin{center}
    \includegraphics[]{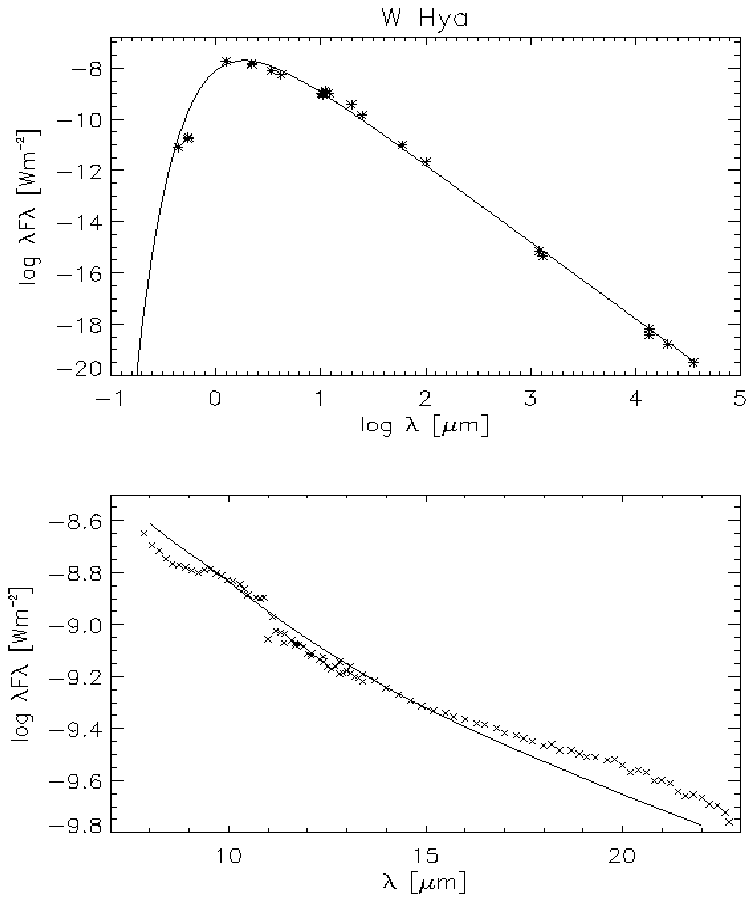}
    \caption{$T_{1} = 1000\,$K, SilDL, $\tau_{0.5} = 0.1$, $T_{\rm{eff}} = 2000\,$K.}
    \label{WHya}
\end{center}
\end{figure}

\begin{figure}
\begin{center}
    \includegraphics[]{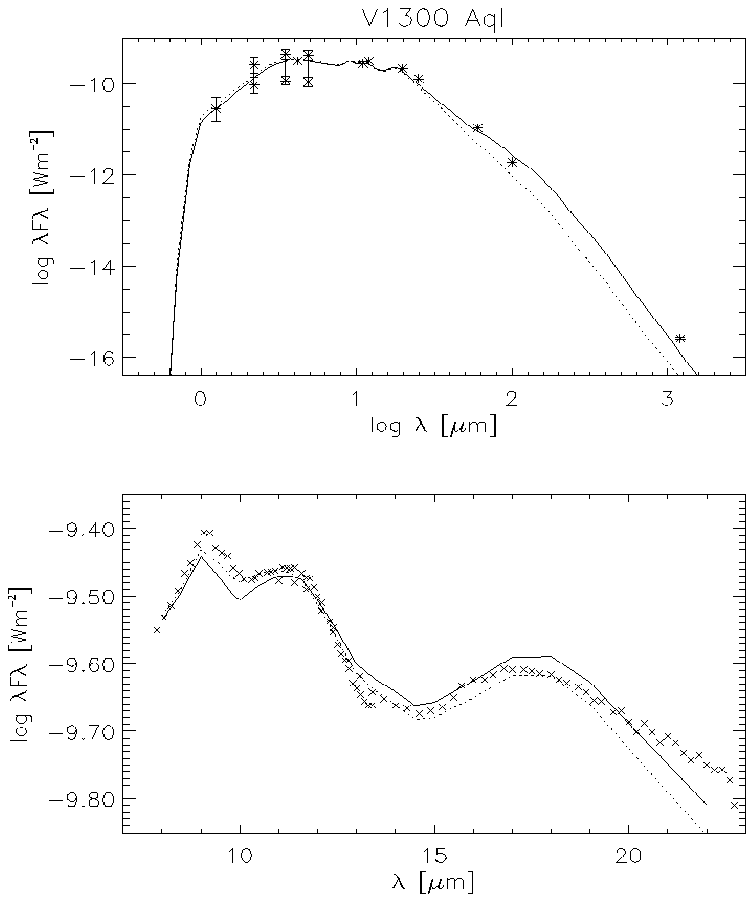}
    \caption{$T_{1} = 1000\,$K, SilOc, $\tau_{0.5} = 50$, $T_{\rm{eff}} = 2500\,$K, $\rho(y) \propto y^{-2}$ between $Y=1$ and $Y=100$,  $\rho(y) \propto y^{-1.5}$ between $Y=100$ and $Y=5000$.}
    \label{oh03620}
\end{center}
\end{figure}

\begin{figure}
\begin{center}
    \includegraphics[]{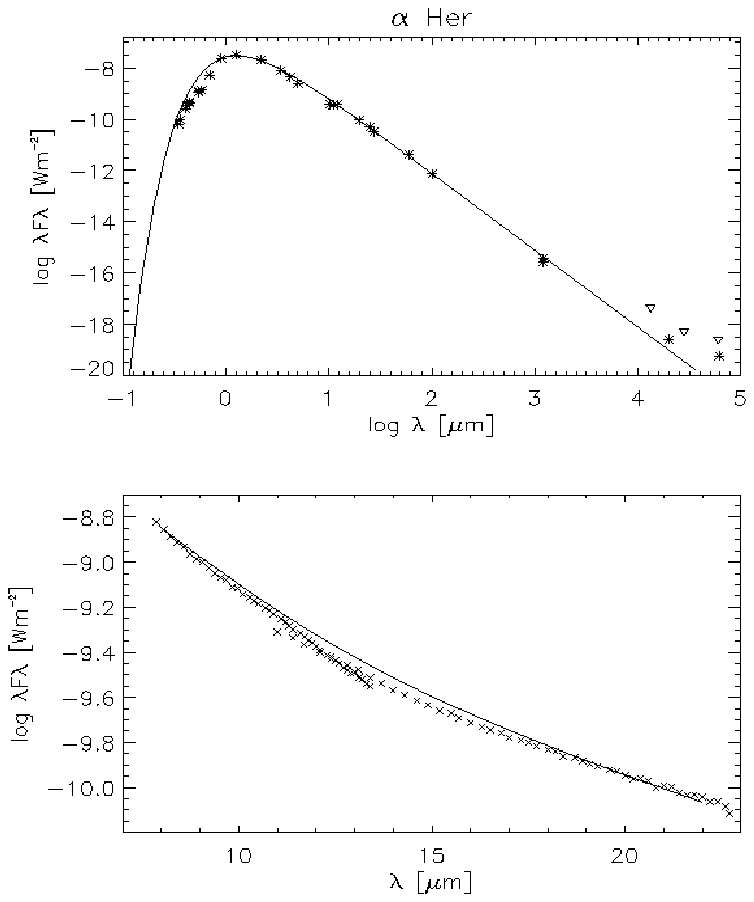}
    \caption{$T_{1} = 1000\,$K, SilDL, $\tau_{0.5} = 0.01$, $T_{\rm{eff}} = 3000\,$K. }
    \label{alfaHer}
\end{center}
\end{figure}

\begin{figure}
\begin{center}
    \includegraphics[]{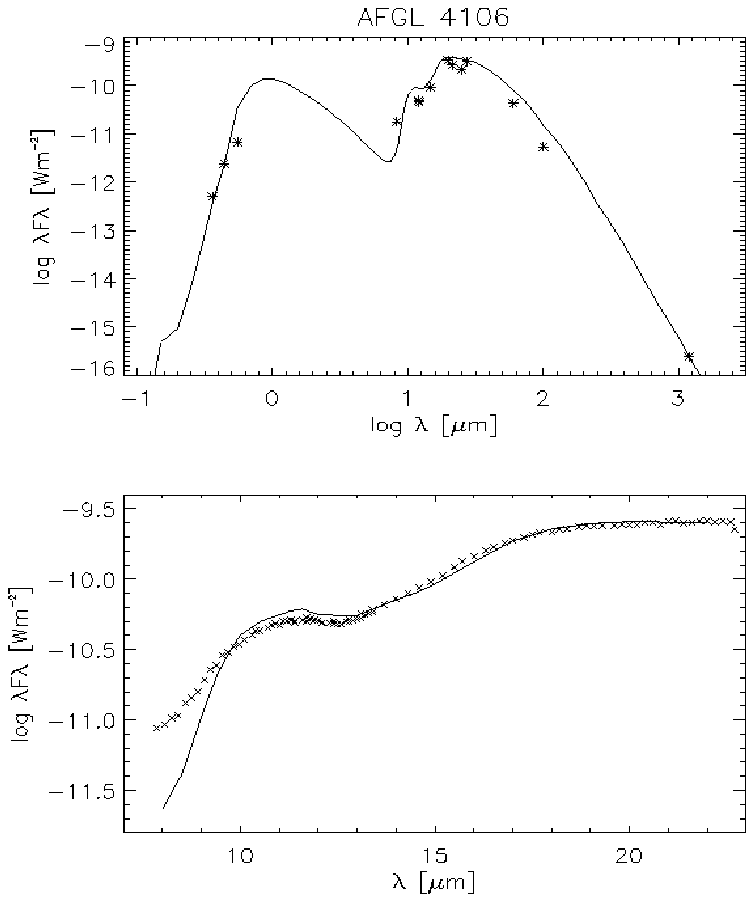}
    \caption{$T_{1} = 150\,$K, SilOw, $\tau_{0.5} = 5$, $T_{\rm{eff}} = 7000\,$K, $Y=10$.}
    \label{afgl4106}
\end{center}
\end{figure}

\begin{figure}
\begin{center}
    \includegraphics[]{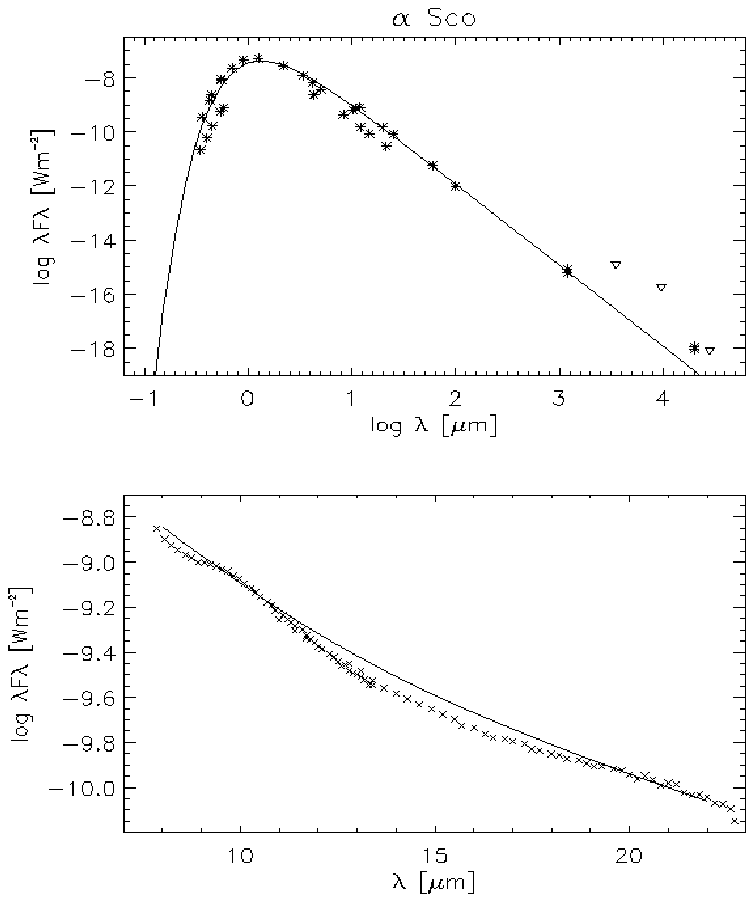}
    \caption{$T_{1} = 1000\,$K, SilDL, $\tau_{0.5} = 0.1$, $T_{\rm{eff}} = 3000\,$K.}
    \label{alfaSco}
\end{center}
\end{figure}

\begin{figure}
\begin{center}
    \includegraphics[]{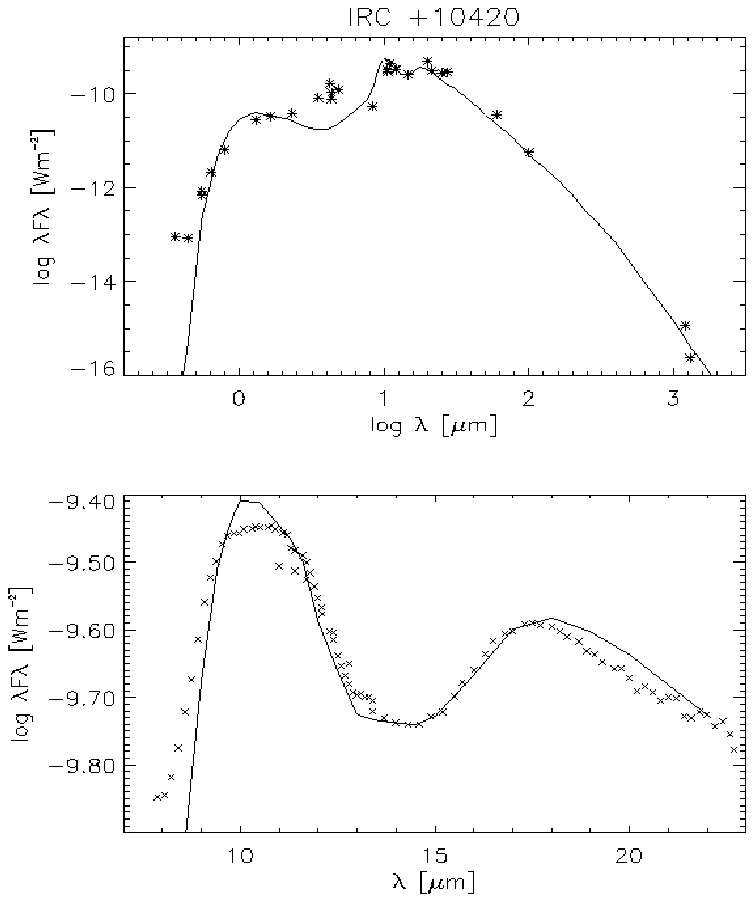}
    \caption{$T_{1} = 400\,$K, SilOw, $\tau_{0.5} = 10$, $T_{\rm{eff}} = 7000\,$K.}
    \label{irc+10420}
\end{center}
\end{figure}

Table \ref{input} gives an overview of the input parameters used for the models, Table \ref{output} does the same for the parameters that were derived from the models. Figures \ref{afgl1922} to \ref{irc+10420} show the SEDs and the LRS spectra together with the best fitting models. The caption of Fig.\,\ref{afgl1922} contains the explanation of the symbols used, for all other figures only the model parameters are given. We comment here on some discrepancies between LRS spectrum and model that have not already been discussed in the article. Several LRS spectra show signs of the atmospheric absorption by HCN and $\textrm{C}_{2}\textrm{H}_{2}$ at $14\,\mu$m. This is the case for IRAS\,$15194$-$5115$, AFGL\,$3068$, IRC\,+$10\,216$ and R\,Scl. At $8\,\mu$m the signature of the SiO fundamental band is visible in the spectra of $o$\,Cet, IK\,Tau, R\,Hya, W\,Hya and $\alpha$\,Sco. In the spectrum of AFGL\,$3068$, SiC goes into self absorption around $12\,\mu$m, which we should be able to model with a larger optical depth. However, a larger value of $\tau_{0.5}$ degrades the quality of the fit to the SED and to the wings of the LRS. This apparent contradiction could indicate a non-spherical mass loss. The discrepancies between LRS spectrum and model for R\,Scl are mainly caused by molecular absorption by HCN at $8\,\mu$m and around $14\,\mu$m in combination with $\textrm{C}_2\textrm{H}_2$. The spectrum of R\,Hya shows a strong molecular band from $\textrm{H}_2\textrm{O}$ to the left of the silicate feature at $9.7\,\mu$m. And the discrepancy between the model and the blue wing of the LRS for AFGL\,$4106$ and IRC\,+$10\,420$ is probably caused by a lack of Fe in the model.
 
\begin{table*}
\centering
\begin{minipage}{165mm}
\caption{This table gives an overview of the input parameters used for the models. $\textrm{A}_v$ is the interstellar reddening derived from the model of \citet{arenou}, $\tau_{0.5}$ is the optical depth at $0.5\,\mu$m, $T_{\rm{eff}}$ is the effective temperature of the central black body and $T_{1}$ is the dust temperature at the inner boundary of the dust envelope. When no error is given for $T_{\rm{eff}}$ or $T_{1}$, this means that the temperature is undetermined. The oxygen rich dust composition is undetermined when $\tau_{0.5}\leq0.1$, except in the case of $\alpha$\,Sco. The last column lists any other parameters that were altered with respect to the default model (see Sect.\,\ref{method}): density structure with parameters $p$ and $y_1$, relative outer radius of the envelope $Y$ or grain size $a$. For Y, only the order of magnitude is determined.}
\label{input}
\begin{tabular}{|l|c|c|c|c|c|l|}
\hline
Name & $\textrm{A}_v$ & $\tau_{0.5}$ & $T_{\rm{eff}}$ & $T_{1}$ & dust composition & other parameters \\ 
     &     	      &              & $\left( K\right)$ & $\left( K\right)$ & & \\ \hline
AFGL\,$1922$ & $0.98\pm0.54$ & $50^{+30}_{-30}$ & $2500$ & $1200^{+500}_{-100}$ & AMC$=93\pm3$\,\% & $p = 1.5\pm0.3$ beyond $y_1 = 100$   \\ 
IRAS\,$15194$-$5115$ & $0.59\pm0.32$ & $15^{+15}_{-5}$ & $2500$  &$900^{+100}_{-100}$  & AMC$=96\pm2$\,\% & $p = 1.5\pm0.3$ beyond $y_1 = 100$  \\ 
AFGL\,$3068$ & $0.15\pm0.15$ & $80^{+5}_{-5}$ & $2500$ & $600^{+100}_{-100}$ & AMC$=93\pm5$\,\% &  \\ 
IRC\,+$10\,216$ & $0.06\pm0.15$ & $15^{+5}_{-5}$ & $2500$ & $600^{+100}_{-100}$ & AMC$=97\pm2$\,\% 	& $a = 0.15\pm0.05$ \\ 
IRC\,+$20\,370$ & $0.49\pm0.20$ & $10^{+5}_{-5}$ & $2500$ & $900^{+200}_{-200}$ & AMC$=90\pm2$\,\% &  \\ 
V\,Hya & $0.16\pm0.15$ & $3^{+2}_{-2}$ & $2500$ & $700^{+200}_{-200}$ & AMC$=93\pm2$\,\% &  \\ 
R\,Scl & $0.00\pm0.15$ & $0.5^{+0.2}_{-0.3}$ & $2500^{+500}_{-500}$ & $1000^{+300}_{-200}$ & AMC$=90\pm5$\,\% & $p = 1\pm0.2$ beyond $y_1 = 100$  \\ 
IRC\,+$20\,326$ & $0.32\pm0.15$ & $20^{+10}_{-10}$ & $2500$ & $700^{+100}_{-200}$ & AMC$=97\pm2$\,\% 	& $p = 3\pm0.4$ beyond $y_1 = 100$ \\ 
o\,Cet & $0.07\pm0.16$ & $1^{+1}_{-0.9}$ & $3000^{+500}_{-750}$ & $1000^{+300}_{-300}$ & SilOc &  \\ 
IK\,Tau & $0.32\pm0.17$ & $20^{+10}_{-5}$ & $2500$ & $1000^{+200}_{-200}$ & SilDL & $a = 0.15\pm0.1$ \\ 
WX\,Psc & $0.12\pm0.17$ & $70^{+10}_{-20}$ & $2500$ & $1000^{+200}_{-100}$ & SilOw & $p = 1.5\pm0.4$ beyond $y_1 = 100$  \\ 
R\,Hya & $0.23\pm0.15$ & $0.1^{+0.4}_{-0.09}$ & $2500^{+500}_{-500}$ & $300^{+100}_{-100}$ & SilDL &  \\ 
W\,Hya & $0.20\pm0.15$ & $0.1^{+0.9}_{-0.09}$ & $2000^{+500}_{-500}$ & $1000$ & SilDL &  \\ 
V$1300$ Aql & $0.46\pm0.26$ & $50^{+10}_{-10}$ & $2500$ & $1000^{+200}_{-100}$ & SilOc & $p = 1.5\pm0.3$ beyond $y_1 = 100$  \\ 
$\alpha$\,Her & $0.28\pm0.17$ & $0.01^{+0.09}_{-0.009}$ & $3000^{+500}_{-500}$ & $1000$ & SilDL &  \\ 
AFGL\,$4106$ & $1.34\pm0.67$  & $5^{+3}_{-2}$ & $7000$ & $150^{+50}_{-50}$ & SilOw & $Y = 10^{+10}_{-5}$ \\ 
$\alpha$\,Sco & $0.76\pm0.40$ & $0.1^{+0.1}_{-0.05}$ & $3000^{+1000}_{-500}$ & $1000^{+400}_{-400}$ & SilDL & \\ 
IRC\,+$10\,420$ & $1.96\pm1.03$ & $10^{+5}_{-5}$ & $7000$ & $400$ & SilOw & \\ \hline	 
\end{tabular}
\end{minipage} 
\end{table*}

\begin{table*}
\centering
\begin{minipage}{110mm}
\caption{This is an overview of the output parameters of the presented models. L is the luminosity, $r_1$ is the distance between the dust envelope and the central star, $r_1/r_c$ is the ratio of this distance to the radius of the central star. The hydrodynamics calculations give a value for the mass-loss rate $\dot M$ and the terminal outflow velocity $\textrm{V}_\textrm{e}$. If the terminal outflow velocity falls below $5\,\textrm{km}\,\textrm{s}^{-1}$, the values of $\dot M$ and $\textrm{V}_\textrm{e}$ can not be trusted (indicated with an exclamation mark), if $\textrm{T}_{\rm{eff}}$ drops below a critical temperature inherent to the model, all values listed in the table are uncertain (indicated with an asterisk).}
\label{output}
\begin{tabular}{|l|c|c|c|c|c|}
\hline
Name		& L 		& $\dot M$   		&  $\textrm{V}_\textrm{e}$	& $r_{1}$		&   $r_{1}/r_{c}$   \\ 
              	&$\left(L_{\odot}\right)$&$\left(M_{\odot}\,\textrm{yr}^{-1}\right)$&$\left(\textrm{km\,s}^{-1}\right)$&$\left(\textrm{cm}\right)$&
		\\ \hline
AFGL\,$1922$    & $1.03\,10^4$  & 		 	& 	& $2.15 \, 10^{14}$ 	& $5.71$ 	\\ 
IRAS\,$15194$-$5115$& $1.17\,10^4$&		 	& 	& $4.07\,10^{14}$ 	& $10.2$ 	\\      
AFGL\,$3068$ $!$& $6.78\,10^3$ 	& $3.34\,10^{-5}$ 	& $4.85$& $9.47\,10^{14}$ 	& $31.1$ 	\\
IRC\,+$10\,216$ 	& $7.87\,10^3$ 	& $1.73\,10^{-5}$ 	& $6.65$& $1.03\,10^{15}$ 	& $31.3$ 	\\
IRC\,+$20\,370$ 	& $1.12\,10^4$ 	& $7.87\,10^{-6}$ 	& $15.1$& $3.90\,10^{14}$ 	& $9.92$ 	\\
V\,Hya 		& $9.78\,10^3$ 	& $4.72\, 10^{-6}$ 	& $13.0$& $6.65 \, 10^{14}$ 	& $18.1$ 	\\
R\,Scl 		& $1.32\,10^4$ 	& 		  	& 	& $2.80 \, 10^{14}$ 	& $6.57$ 	\\
IRC\,+$20\,326$ & $8.82\,10^3$ 	&  			& 	& $6.84 \,10^{14}$ 	& $19.6$ 	\\
o\,Cet 		& $1.06\,10^4$ 	& $1.15\,10^{-6}$ 	& $16.2$& $2.03\,10^{14}$ 	& $7.63$ 	\\
IK\,Tau $^*$ 	& $1.06\,10^4$ 	& $8.23\,10^{-6}$ 	& $14.1$& $2.38 \,10^{14}$ 	& $6.26$ 	\\
WX\,Psc 	& $7.38\,10^3$ 	& 			& 	& $2.35\,10^{14}$ 	& $7.36$ 	\\ 
R\,Hya $!$ 	& $1.03\,10^4$ 	& $3.57 \, 10^{-7}$ 	& $3.05$& $1.07 \, 10^{15}$ 	& $28.2$ 	\\
W\,Hya $^*$ 	& $1.26\,10^4$ 	& $1.07\,10^{-7}$ 	& $7.11$& $1.37\,10^{14}$ 	& $2.10$ 	\\
V$1300$\,Aql 	& $9.51\,10^3$ 	& 		 	& 	& $2.34\,10^{14}$ 	& $6.46$ 	\\
$\alpha$\,Her 	& $2.04\,10^4$ 	& $5.09\,10^{-8}$ 	& $7.16$& $2.16\,10^{14}$ 	& $5.86$ 	\\
AFGL\,$4106$ $!$& $2.48\,10^5$ 	& $9.50\,10^{-4}$ 	& $3.73$& $1.81\,10^{17}$ 	& $7.68\,10^{3}$\\
$\alpha$\,Sco 	& $7.98\,10^4$ 	& $1.41\,10^{-7}$ 	& $10.1$& $4.27\,10^{14}$ 	& $5.86$ 	\\
IRC\,+$10\,420$ 	& $2.88\,10^5$ 	& $4.78\,10^{-4}$ 	& $10.0$& $2.21\,10^{16}$ 	& $8.71\,10^{2}$\\ \hline 
\end{tabular}
\end{minipage} 
\end{table*}

\clearpage

\label{lastpage}
\end{document}